\documentclass[aps,twocolumn,groupedaddress,superscriptaddress]{revtex4-1}
\usepackage[utf8]{inputenc}
\usepackage[pdftex]{graphicx}
\usepackage{amsmath}
\usepackage{amsfonts}
\usepackage{times}
\usepackage{soul}
\usepackage[colorlinks,citecolor=blue]{hyperref}
\usepackage[colorinlistoftodos, color=green!40, prependcaption]{todonotes}

\newcommand{\gcc}{g$\,$cm$^{-3}$}

\begin{document}

\title{A Model of Ramp Compression of Diamond from Ab Initio Simulations}
\author{F. Gonz\'alez-Cataldo}
\affiliation{Department of Earth and Planetary Science, University of California, Berkeley, California 94720, USA}
\author{B.K. Godwal}
\affiliation{Department of Earth and Planetary Science, University of California, Berkeley, California 94720, USA}
\author{K. Driver}
\affiliation{Department of Earth and Planetary Science, University of California, Berkeley, California 94720, USA}
\affiliation{Lawrence Livermore National Laboratory, Livermore, CA 94550, USA}
\author{R. Jeanloz}
\affiliation{Department of Earth and Planetary Science, University of California, Berkeley, California 94720, USA}
\affiliation{Department of Astronomy, University of California, Berkeley, California, USA}
\affiliation{Miller Institute for Basic Research in Science, University of California, Berkeley, CA 94720, USA}
\author{B. Militzer}
\affiliation{Department of Earth and Planetary Science, University of California, Berkeley, California 94720, USA}
\affiliation{Department of Astronomy, University of California, Berkeley, California, USA}
\date{\today}

\begin{abstract}
Ramp compression experiments characterize high-pressure states of matter at temperatures well below those present in shock compression. However, because temperature is typically not directly measured during ramp compression, it is uncertain how much heating occurs under these shock-free conditions. Here, we performed a series of ab initio simulations on carbon in order to match the density-stress measurements of Smith et al.~[Smith, \emph{et al.}, Nature \textbf{511}, 330 (2014)]. We considered isotropically as well as uniaxially compressed solid carbon in the diamond and BC8 phases, with and without defects, as well as liquid carbon. Our idealized model ascribes heating during ramp compression to an initially uniaxially compressed cell transforming isochorically into an isotropically (hydrostatic equivalent) compressed state having lower internal energy, hence higher temperature so as to conserve energy. Multiple such heating events can occur during a single ramp experiment, leading to higher temperatures than with isentropic compression.
Comparison with experiments shows that heating alone does not explain the equation of state measurements on diamond, instead implying that a significant uniaxial stress component remains present at high compression.
The temperature predictions of our ramp compression model remain to be verified with future laboratory measurements.
\end{abstract}


\maketitle

\section{Introduction}
High-pressure experiments and theoretical investigations are essential to understanding the extreme conditions present within planetary interiors, because the equations of state of constituent materials are not sufficiently well characterized at present. Static compression experiments are able to reach such extreme conditions by compressing matter using diamond anvil cells (DAC), where sample temperatures can be controlled by laser heating, but the applied hydrostatic pressures are limited to $\sim$6-9 Mbar~\cite{Tateno2012,Anzellini2013,Hemley1997}. On the other hand, in dynamic shock compression, an impactor or a high-intensity laser pulse generates a shock wave that propagates through the sample, achieving higher pressures than static compression but with significant heating. Shock temperatures are  difficult to measure accurately, resulting in material properties with poorly characterized temperature dependencies~\cite{Hicks2008,Brygoo2007,Nagao2006}. Dynamic shocks have an inherent rise time due to dissipative processes, such as viscosity or scattering, that cause an increase in entropy~\cite{Swift2008}. The sample usually melts upon reaching Mbar pressures, which limits the ability to characterize crystalline solids at high pressure with dynamic shock compression experiments.

One promising technique to solve the quandary of dynamically probing high densities at lower temperatures is ramp compression~\cite{Swift2008,Bradley2009,Smith2014,Smith2018}, which utilizes a continuously increased laser power source in order to avoid the formation of shocks that can substantially heat the material during compression.
The idea is to design the shape of the ramp pulse in a way that either no shock is formed or the shock forms as late as possible, maintaining a relatively large portion of the material under a pure ramp as opposed to a shock~\cite{Swift2008}.
This is an intrinsically unstable situation, as a finite-amplitude stress wave tends to steepen into a shock as it moves forward.
One typically assumes that ramp experiments compress materials along a thermodynamic path that is close to an isentrope~\cite{Swift2008,Smith2018}, resulting in significantly less heating and much higher densities than shock waves~\cite{Hansen2021}. Access to this dense, low temperature regime is key to understanding the interiors of planets, with central pressures reaching 100~Mbar~\cite{Ross1981,Hubbard1981,Benedetti1999,Guillot2005,Valencia2009,Seager2007,Wahl2021}. Since they primarily cool through convection, their interior temperature profiles are assumed to be isentropic, thus, ramp compression is well suited to probe conditions in the interiors of super-Earths and giant planets, as well as to characterizing novel high-pressure crystal structures, therby testing the validity of theoretical predictions~\cite{Oganov2007,Pickard2009,Wilson2014,Gonzalez-Cataldo2014,Gonzalez-Cataldo2016,Wu2021}.

Recent efforts in developing the ramp-compression technique have been successful in probing relatively cold and solid materials, such as diamond, at ultra-high pressures~\cite{Swift2008,Bradley2009,Smith2014,Smith2018,Ross1981,Hubbard1981,Benedetti1999,Guillot2005,Valencia2009,Seager2007,Oganov2007,Pickard2009,Wilson2014,Benedict2014,Wang2013,Ping2013,Wang2014,Eggert2015,Wang2016,Wei2017,Han2021,Coppari2021}. While the experiments provide accurate measurements of stress and density, similarly accurate measurements of temperature are lacking~\cite{Ping2013,Coppari2021}. In order to predict a material's response to compression, one needs to reliably know its equation of state that describes how the internal energy and pressure depend on density and temperature. Currently, there is no theoretical framework, equivalent to the Rankine-Hugoniot equations for shock compression~\cite{Zel2002}, that provides a complete description of the ramp compression. There have been some preliminary temperature estimates, based on approximations of the plastic work done by shocks~\cite{Bradley2009,Ping2013,Fratanduono2020}, but such methods lack a microscopic formalism. 

We present two alternative interpretations of existing measurements on diamond ramp compression: i) if the measured stresses in the sample are isotropic (fluid-like response, as found for shock-compression measurements well above the Hugoniot elastic limit), then the temperatures must be high – well above the melting temperature – in order to match the observed strains (densities); ii) alternatively, if the stresses are nonisotropic, implying the presence of strength under ramp loading, then the observed strains are consistent with much lower temperatures, and the sample remains well below the melting temperature for the ramp loading we consider.  The second interpretation is consistent with previous estimates of the temperature under ramp loading, and with the lack of any empirical evidence of melting in the experiments we consider, as discussed below. 

In this manuscript, we present a simplified model, based on thermodynamics and density functional theory molecular dynamics (DFT-MD) simulations~\cite{Hohenberg1964,Kohn1965}, to estimate the temperature increase along a ramp-loading path.
We provide a mechanism of energy transfer that explains how temperature increases during the loading process in terms of a series of uniaxial compression steps, as an analogue to the actual, ramp compression of solid materials.

We apply our model to diamond and show that, when many compression steps are employed, the resulting compression paths are compatible with predictions from a multi-shock Hugoniot curve approach, a procedure often used to approximate an isentrope by breaking up a single-shock into multiple, smaller successive shocks~\cite{Driver2018}.  We show that the temperatures along the ramp compression path are much higher than an isentropic profile.

The paper is organized as follows: Sec.~\ref{sec:SimDetails} provides details of our simulations methods. Sec.~\ref{sec:Motivation} discusses the experimental results by Smith et al. that motivated this work and introduce six different hypotheses to match the density-stress data points with first-principles simulations. Elastic constants and stability criteria are analyzed. Sec.~\ref{sec:PlasticWork} discusses a model based on plastic work to estimate temperature in ramp-compression model. Sec.~\ref{sec:RWCmodel} introduces our ramp wave compression model. Sec.~\ref{sec:Multishock} compares the similarities between our model and a multi-shock Hugoniot approach. Finally, we present our conclusions in Sec.~\ref{sec:Conclusions}.

\section{Simulation details}\label{sec:SimDetails}

Our DFT-MD simulations were performed in cubic supercells of 64 atoms for hydrostatically compressed diamond, while uniaxial compressions were performed in tetragonal supercells of 96 atoms. For the BC8 crystal structure, we used cubic supercells with 128 atoms and tetragonal supercells with 192. Simulations of carbon in the simple cubic (SC) phase were performed in a 64 atoms supercells, while the uniaxial compressions of this phase were done in tetragonal 128 atoms supercell.
Previous studies~\cite{Militzer2016} demonstrate that these sizes are sufficient to achieve convergence for both pressure and energy in DFT-MD simulations of diamond. We also include a study of finite-size effects in our supplementary material.

We also generated BC8 supercells with defects, which were generated by removing atoms from an ideal BC8 supercell in consecutive steps. We start with one defect. For the next and all following steps, we remove the atom (or one of the atoms) that is farthest away from all the previously generated defects. This procedure provides us with a reasonable starting configuration for the MD simulations with defects. As we will discuss later, these defects will migrate at elevated temperatures rendering the initial configuration less important.

The temperature was regulated with a Nosé thermostat~\cite{Nose1984,Hoover1985} in the NVT ensemble. We sampled the Brillouin zone at the  $\Gamma$-point in all our supercells, except for the SC phase, where we used the Balderesci point. We used a time step of 1 fs and a total simulation time of at least 5 ps, which is enough to yield sufficiently converged values of the energy and pressure. The calculations were performed under the projector augmented-wave (PAW)~\cite{PAW} method as implemented in VASP~\cite{Kresse1999}, using the generalized-gradient approximation (GGA) of Perdew, Burke and Ernzerhof (GGA-PBE)~\cite{PBE}. The 2s$^2$2p$^2$ electrons were taken in their valence configuration of the pseudopotential, and an energy cutoff of 900~eV was used for the plane-wave expansion.

\section{Ramp Compression Experiments of Diamond and Equation of State}\label{sec:Motivation}

\begin{figure*}[!hbt]
    \includegraphics[width=8cm]{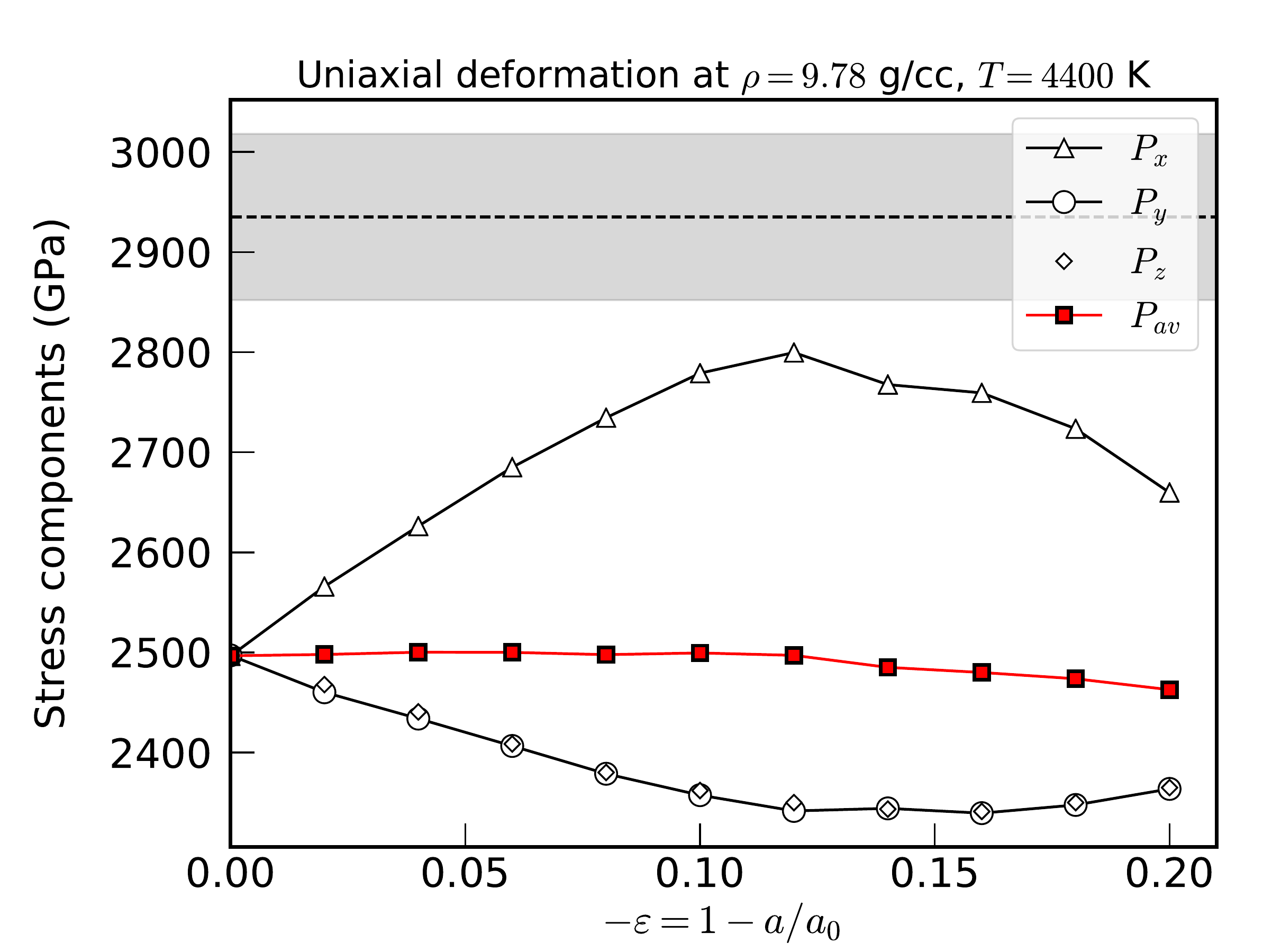}
    \includegraphics[width=8cm]{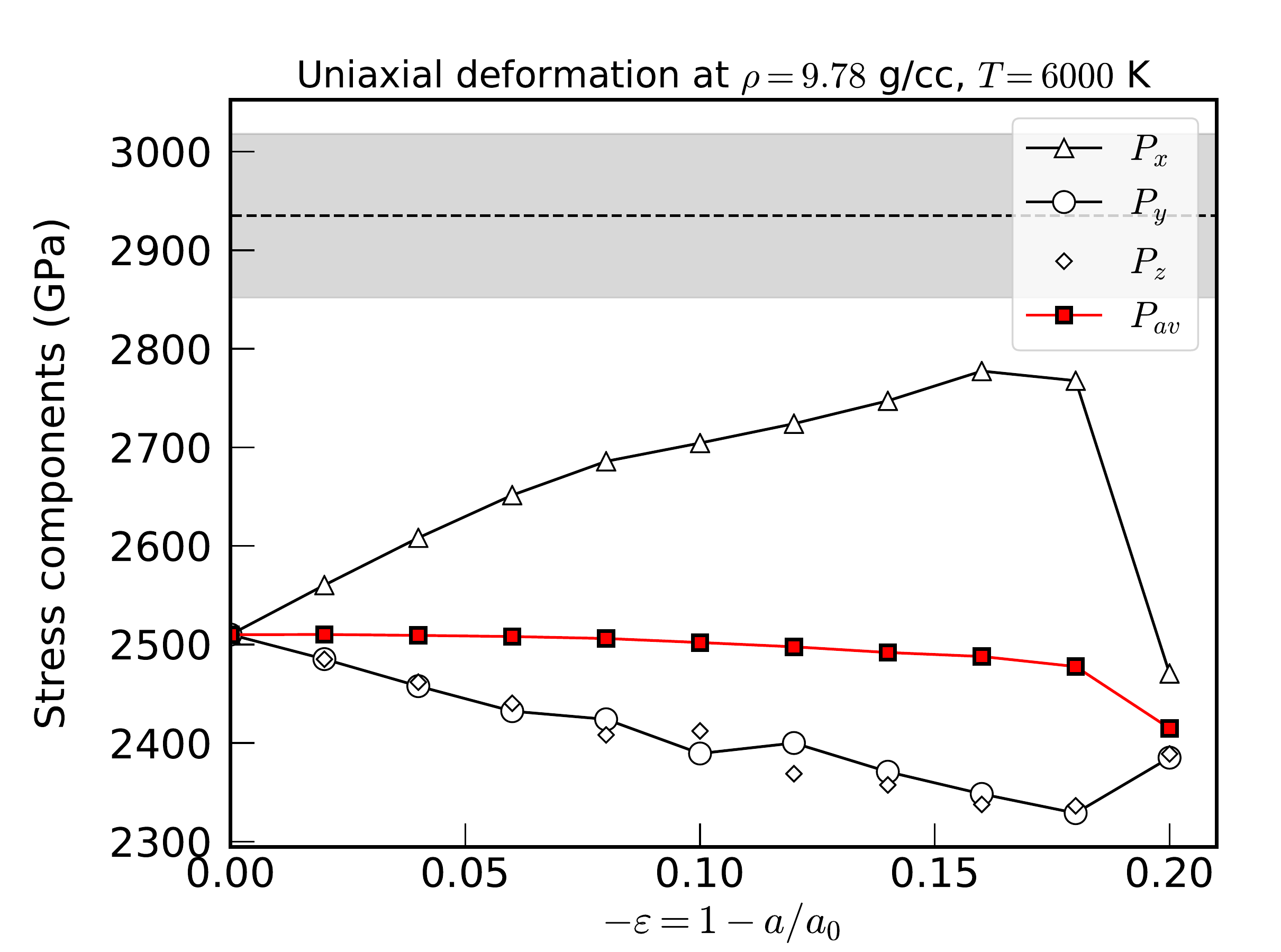}
    \includegraphics[width=8cm]{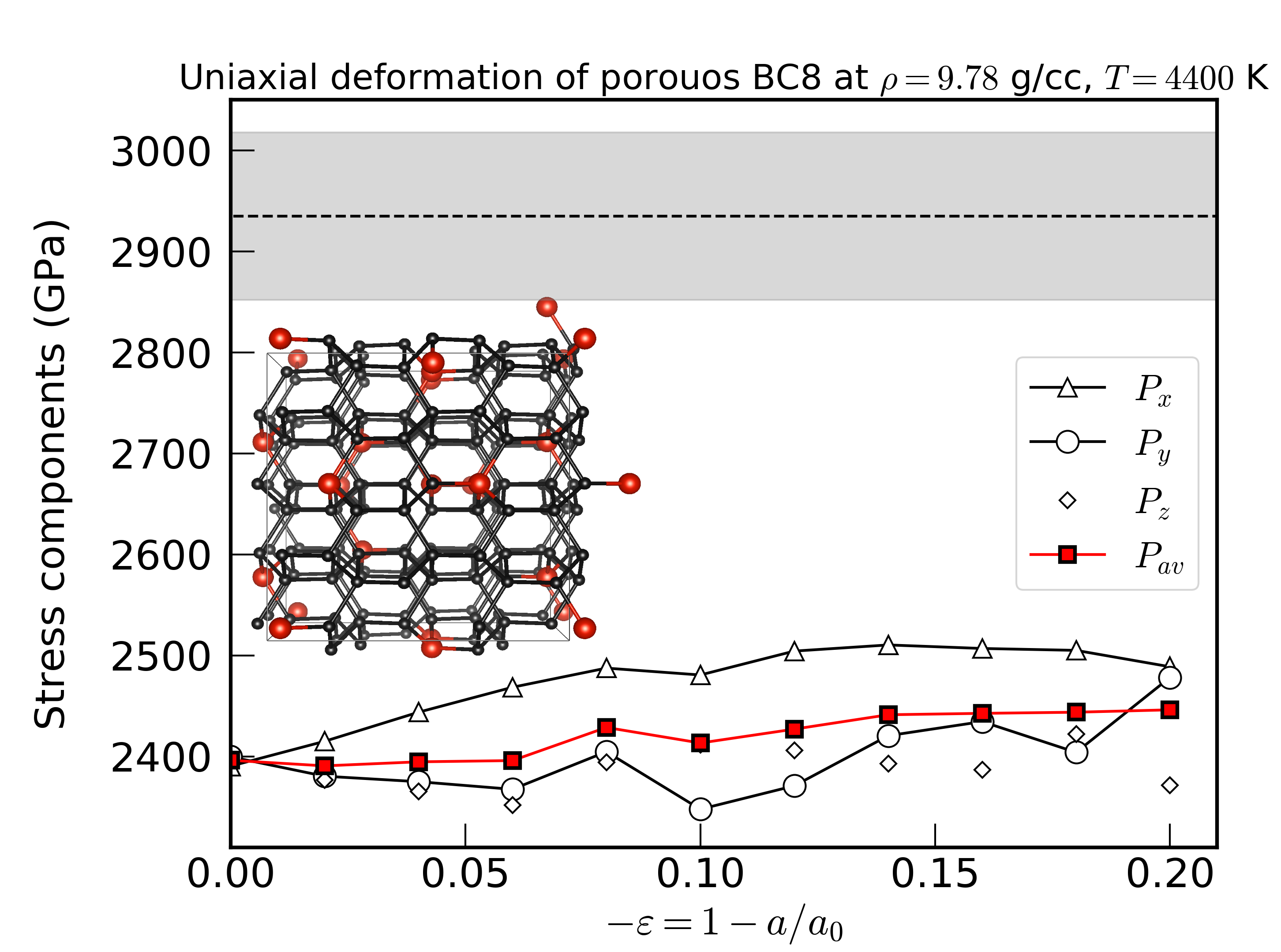}
    \includegraphics[width=8cm]{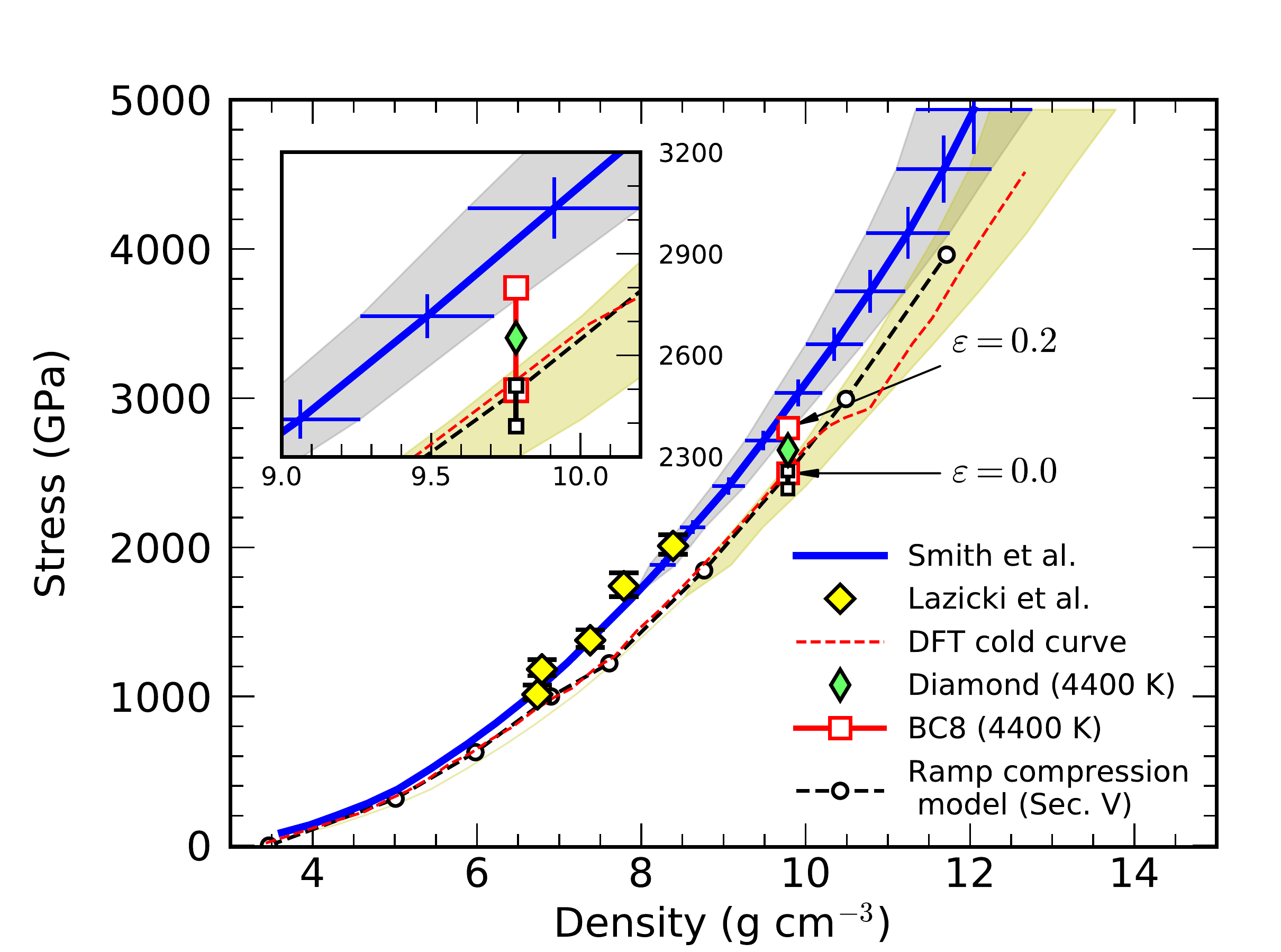}
    \caption{Normal stress components vs. strain derived from DFT-MD simulations for uniaxially strained BC8 supercells of carbon at 9.78 \gcc. The shaded grey area in the first three panels represents the error bars in the stress reported at this density (2935 GPa) by Smith et al.~\cite{Smith2014}. The two upper panels result from defect-free simulations at 4400 and 6000 K. The lower left panel depicts predictions from a BC8 cell with 7.8\% porosity and 4400 K, where the vacancies are represented by red spheres. The lower right panel compares the normal stresses reported by Smith et al.~\cite{Smith2014} and Lazicki et al.~\cite{Lazicki2021} with our values of the stress in the compression direction, $P_x$, where the red and black squares connected by a line in this panel represent the entire range of $P_x$ values for uniaxial compressions at 4400 K for BC8 cells with and without defects, respectively. These correspond to uniaxial compressions up to $\varepsilon=$20\% change in lattice constant. The green, thin diamond corresponds to the pressure of carbon in the diamond structure at the same density, 9.78~\gcc, and 4400~K. The yellow shaded area shows the Smith \emph{et al.} data with uncertainties after the densities have been rescaled by a factor of 1.08 (see discussion in main text). The dashed black line with circles corresponds to our proposed ramp compression model, discussed in section~\ref{sec:RWCmodel}.} 
    \label{fig:porous}
\end{figure*}

The starting point for our theoretical investigation are the experiments of Bradley et al.~\cite{Bradley2009}, Lazicki et al.~\cite{Lazicki2021}, and Smith et al.~\cite{Smith2014}, who ramp-compressed diamond to unprecedented stress conditions of 800, 2000, and 5000 GPa, respectively. The goal of this paper is to investigate the possible thermodynamic conditions along these ramp compression paths. Matching the Smith et al. data with DFT-MD simulations proved to be difficult because for a given density, 
our DFT-MD simulations predict normal stresses that are $\sim$8-12\% lower than what was experimentally determined
(see~Fig.~\ref{fig:porous}). We try to interpret this discrepancy by exploring a number of different hypotheses: (i) First, we performed DFT-MD simulations of carbon at pressures beyond 1 TPa in the BC8 structure at different temperatures, as this is the phase predicted to be most thermodynamically stable under these conditions. (ii) We perform simulations of the diamond structure. (iii) We assume the sample has melted and match the stress-density measurements with liquid simulations. (iv) Then we study whether uniaxially strained BC8 solids can match the experimental data. (v) We performed simulations of BC8 cyrstals with defects because they lower the density. (vi) Finally we consider a density-rescaling argument to take into account the initial density of the sample used in the experiments.

(i) With DFT-MD simulations, we first aim to reproduce the density-stress values of Smith et al.~\cite{Smith2014} by assuming a solid structure that is in hydrostatic equilibrium. We focused on a density of 9.78 \gcc\, and performed simulations for the BC8 structure under hydrostatic conditions (cubic cells with $P=P_x=P_y=P_z$) for two temperatures, 4400 and 6000 K, which are close to the melting temperature at this density~\cite{Benedict2014,Nguyen-Cong2020,Willman2020}. We expect them to yield the highest pressures for this density. For these two temperatures, we obtained pressures of 2496 and 2510 GPa respectively at this density ($\varepsilon=0$ in the upper panels of Fig.~\ref{fig:porous}), which are 16\% and 14\% below the experimentally reported stress value of 2935$\pm$83 GPa for this density (see bottom right panel of Fig.~\ref{fig:porous}). When the uncertainty of the measured density ($\sim$2.7\%) is included, $\chi^2$ deviations
\footnote{We define $\chi^2 = \left(\frac{\rho_{\rm model} - \rho_{\rm exp}} {\delta \rho_{\rm exp}} \right)^2 + \left(\frac{P_{x,\rm model}-P_{x,\rm exp}} {\delta P_{x,\rm exp}}\right)^2$ }
of 8.0 and 6.5 are obtained, respectively, when the density-stress measurements are compared with the simulation results for the two temperatures. These discrepancies cannot be bridged by increasing the simulation temperature further because thermal pressure is small compared to the static pressure and--more importantly--carbon is predicted to melt at $\sim$6000 K for these pressures~\cite{Benedict2014}. 

\begin{figure}[!hbt]
\includegraphics[width=8cm]{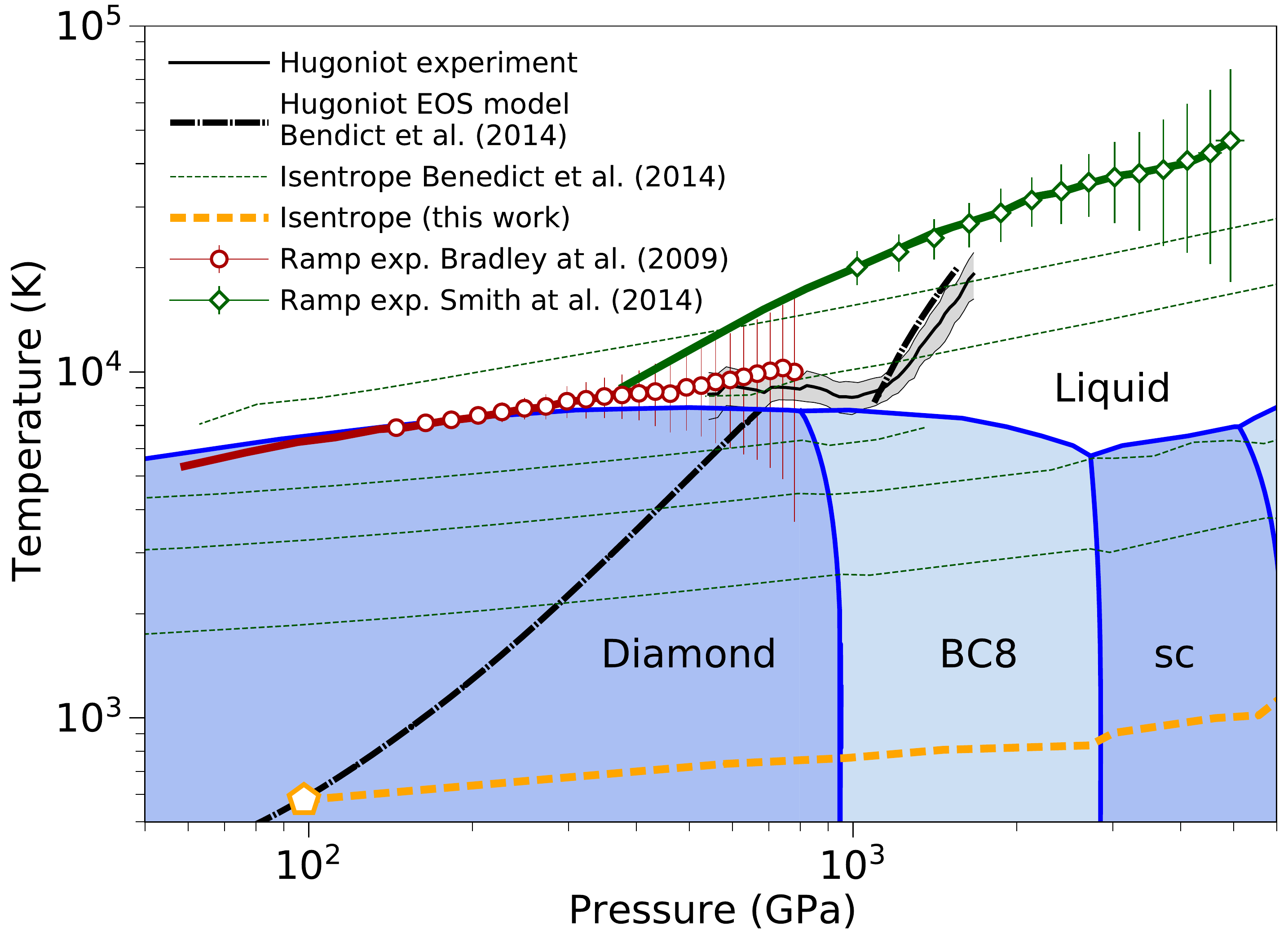}
\caption{Phase diagram of carbon. DFT-MD simulations of dense, liquid carbon~\cite{Benedict2014} were used to attribute a hypothetical temperature to the density-stress data values reported from Bradley et al.~\cite{Bradley2009} and Smith et al.'s~\cite{Smith2014} ramp compression experiments. The vertical lines represent the temperature uncertainties that were obtained by propagating the experimental error bars. The orange dashed curve represents the isentrope that crosses the principal Hugoniot curve at 110 GPa, the diamond elastic limit.}
    \label{fig:PhaseDiagram}
\end{figure}

(ii) The recent work by Lazicki et al.~\cite{Lazicki2021} provided stress-density measurements up to 2000 GPa that are in fairly good agreement with the results of Smith et al. (see yellow diamonds in Fig.~\ref{fig:porous}). In addition, (111) and (220) x-ray diffraction lines demonstrated that the sample remained in the diamond crystal structure up to $\sim$2000 GPa, even though the BC8 phase is predicted to be thermodynamically more stable beyond 950 GPa (see Fig.~\ref{fig:PhaseDiagram}). We thus performed simulations of the diamond structure at 4400~K and 9.78~\gcc\, and obtained a pressure of 2652$\pm$2 GPa. This is 6\% higher than the pressure of the BC8 phase but still not high enough to match the stress reported in the experiments.

(iii) As we were not able to match the density-stress data points of Smith et al. with a solid structure under hydrostatic conditions, we investigated whether these measurements could be matched with simulations of liquid carbon, even though the experiments had shown no signatures of melting along the ramp-compression path. We used our EOS for liquid carbon~\cite{Benedict2014} to reproduce the density-stress conditions of 9.78$\pm$0.27 \gcc\, and 2935$\pm$83 GPa reported in the experiment. This results in an unexpectedly high temperature of 41900$\pm$5300 K. Still, we were able match all the density-stress conditions reported in Ref.~\cite{Smith2014}, and attribute a temperature to every data point. We propagated the experimental stress and density error bars and plot the inferred temperature conditions in Fig.~\ref{fig:PhaseDiagram}. For the density points 8.24$\pm$0.16 and 12.4$\pm$0.7 \gcc, we infer temperatures of 34700$\pm$3600 K and 47000$\pm$26000 K, respectively. The assumption, that the reported density-stress data points correspond to liquid carbon implies a significant amount of heating has occurred during ramp compression~\cite{Smith2014}. This provided us with motivation to construct a thermodynamic model to estimate the heating in ramp-compression experiments, which we present in Sec.~\ref{sec:RWCmodel}. 

If one assumes the compression is purely isentropic, the temperature does not exceed 1000~K within pressures of 3000~GPa as the orange curve in Fig.~\ref{fig:PhaseDiagram} shows. We used the quasi-harmonic approximation (QHA) to obtain isentropes from the free energies of each solid phase. In particular, we obtain the isentrope that crosses the principal Hugoniot curve at 110 GPa, the pressure reported by Smith et al. as the initial state for their ramp loading. Under the assumption that the compression was isentropic, and no plastic work contributed to the heating, one would be forced to conclude that no melting has occurred in the experiments of Smith et al. On the other hand, one cannot match the reported stress-density data points with simulations of solids under hydrostatic conditions.

(iv) Under this hypothesis, we explore whether the high stresses observed in the experiments are consistent with a solid sample under a high degree of uniaxial strain. It is our goal to match the measured density-stress points with DFT-MD with simulations of strained BC8 and to derive a temperature estimate.


We simulated uniaxially compressed BC8 supercells with the strain applied in the x direction in order to obtain higher $P_x$ values than those derived from hydrostatic simulations at the same density. We adjusted the shape of the simulation cell such that the density was kept constant at 9.78 \gcc\, during the uniaxial compression, the same density we used for hydrostatic compression. In the two upper panels of Fig.~\ref{fig:porous}, we plot the computed normal stresses $P_x$ and $P_y=P_z$ in the uniaxially compressed BC8 cell as a function of x-strain at this density for temperatures of 4400 K and 6000 K.
In the elastic regime, where strain values are small, the stresses increase linearly and $P_x>P_y=P_z$, as expected. For the two temperatures, $P_x$ increases with strain until it reaches maximum values of 2777 and 2800 GPa, respectively, for strain magnitudes of 12 and 16\%. If the BC8 cell is strained further, $P_x$ decreases because mechanical instabilities take place. For very large strain values of 18 and 20\%, the structure becomes unstable and the uniaxially strained BC8 crystal collapses.

Uniaxial compression at temperatures of 4400 and 6000 K leads to maximum stress values of 2777 and 2800 GPa, which are still 5.4\% and 4.6\% lower than the measured stress value of 2935$\pm$83 GPa (see Fig.~\ref{fig:porous}, upper panels). When the uncertainty of the measured density is considered, $\chi^2$ values of 0.7 and 0.5 are obtained that are much lower than those obtained with hydrostatic conditions. There are two ways to interpret the remaining discrepancy. One could argue that $\chi^2$ values less than 1.0 represent satisfactory agreement between simulations and experiments, and the ramp-compressed carbon crystals were strained close to the limit of their structural stability. Alternatively, one could argue that the sample melted in Smith et al.'s experiments, but this would require a significant amount of heating and would not be consistent with x-ray diffraction peaks observed at lower pressures~\cite{Lazicki2021}.
 \begin{figure}[!hbt]
    \includegraphics[width=8cm]{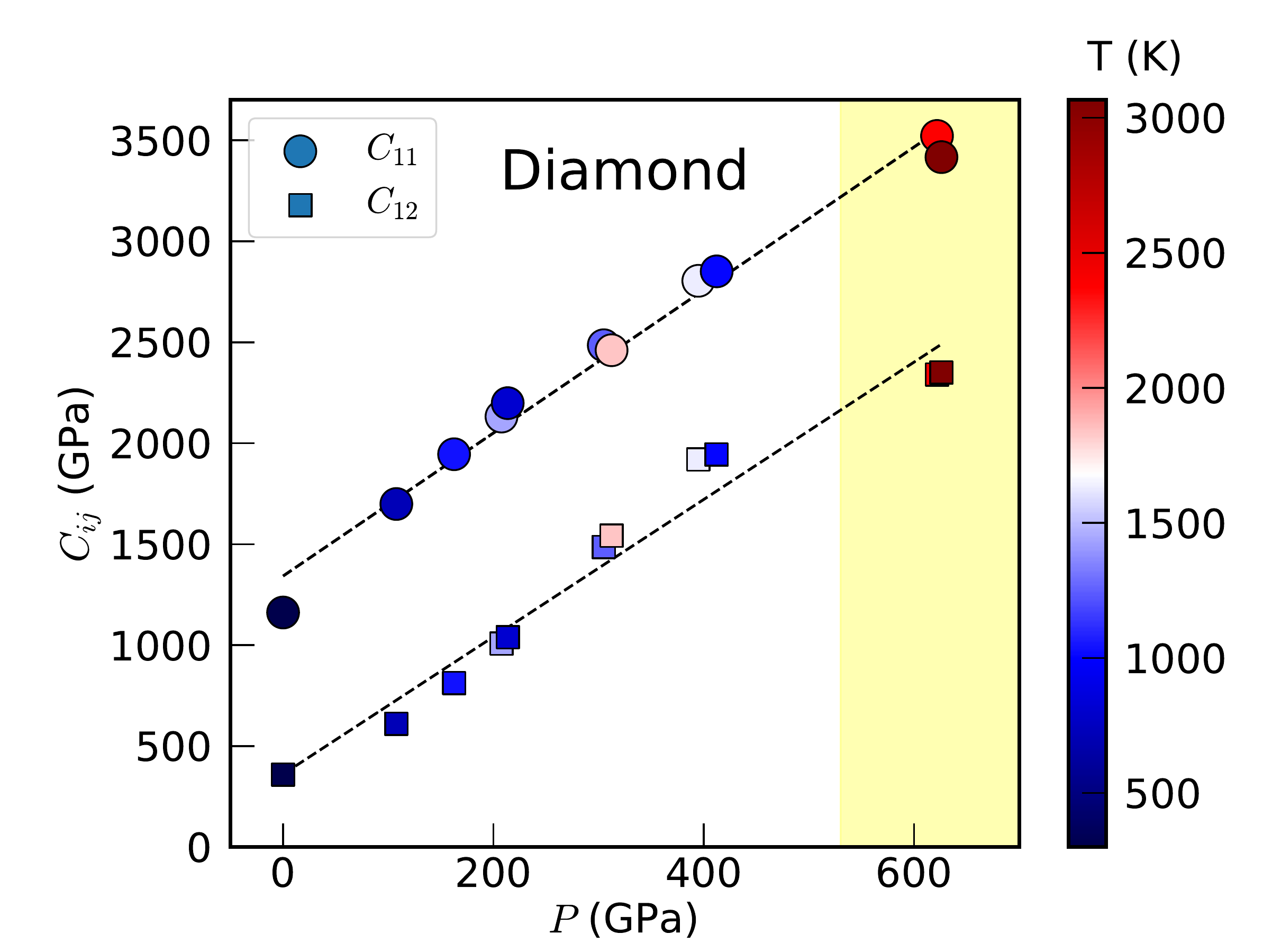}
    \includegraphics[width=8cm]{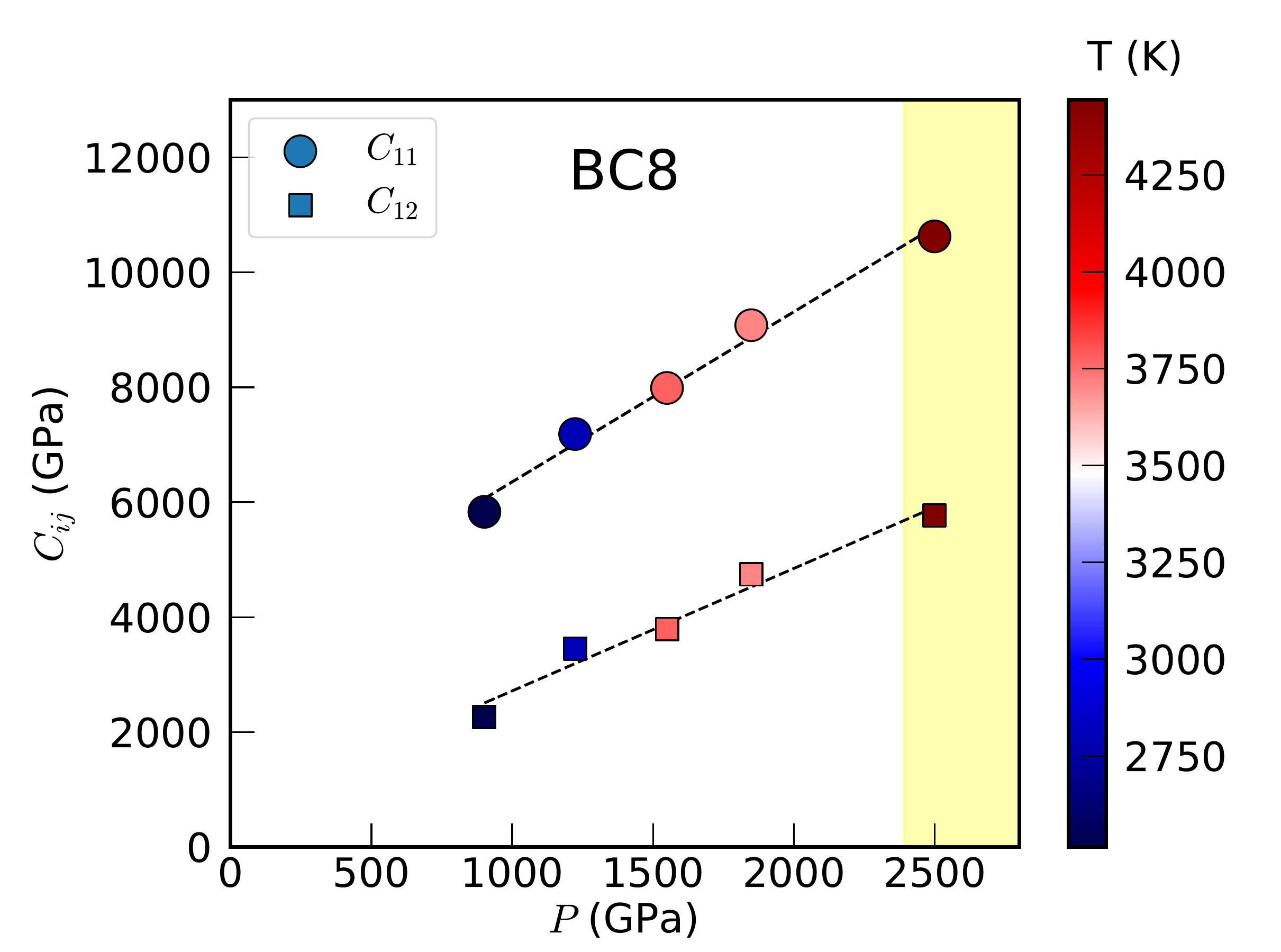}
    \caption{Elastic constants $C_{11}$ and $C_{21}$ of diamond and BC8 carbon as a function of pressure at different temperatures. The shaded region indicates the instability regime, where the Born criterion~\cite{Wang1993} , $(C_{11}-C_{12})/2>P$, is no longer satisfied.}
    \label{fig:Cij}
\end{figure}

\begin{figure}[!hbt]
    \includegraphics[width=8cm]{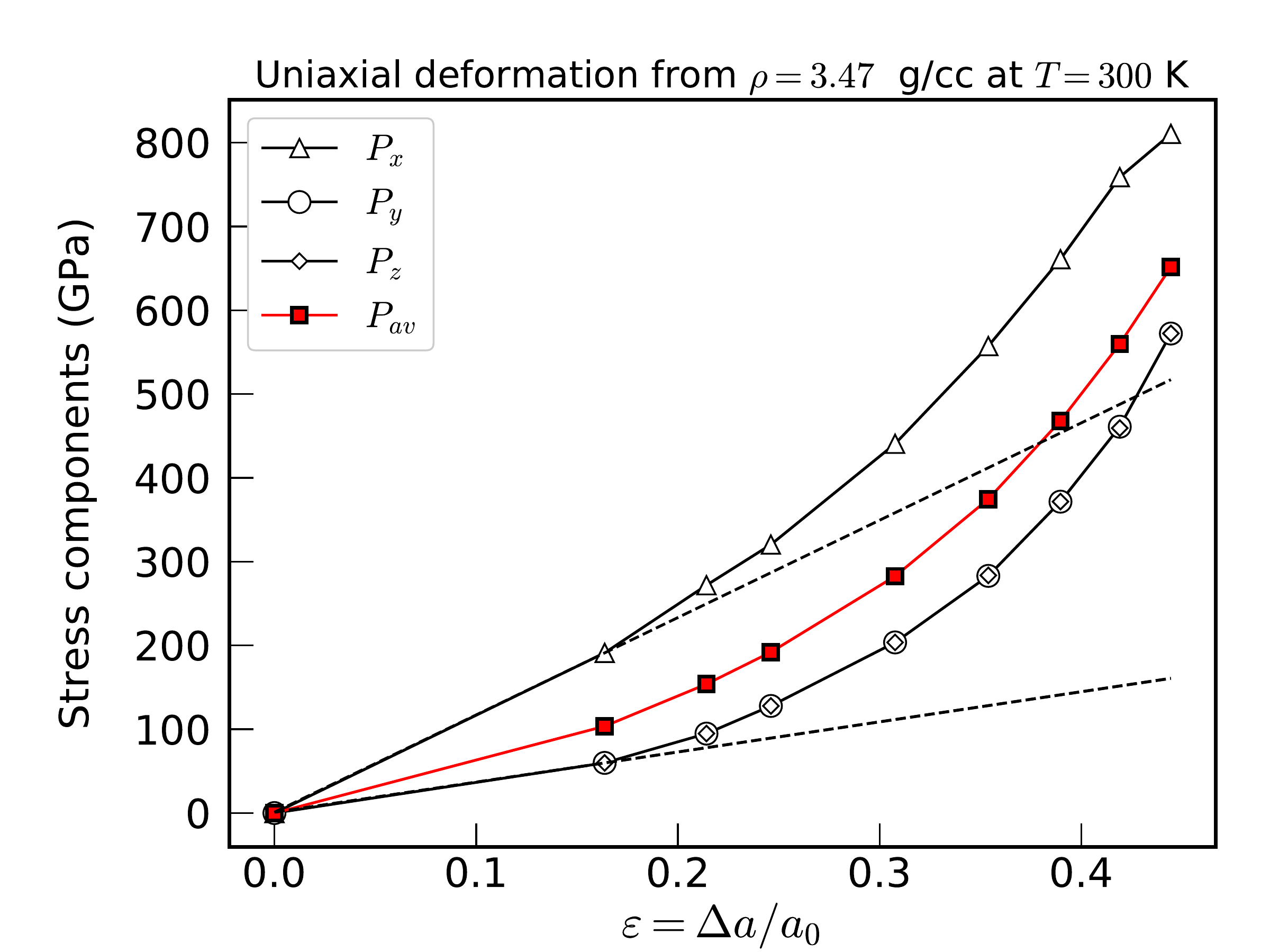}
    \includegraphics[width=8cm]{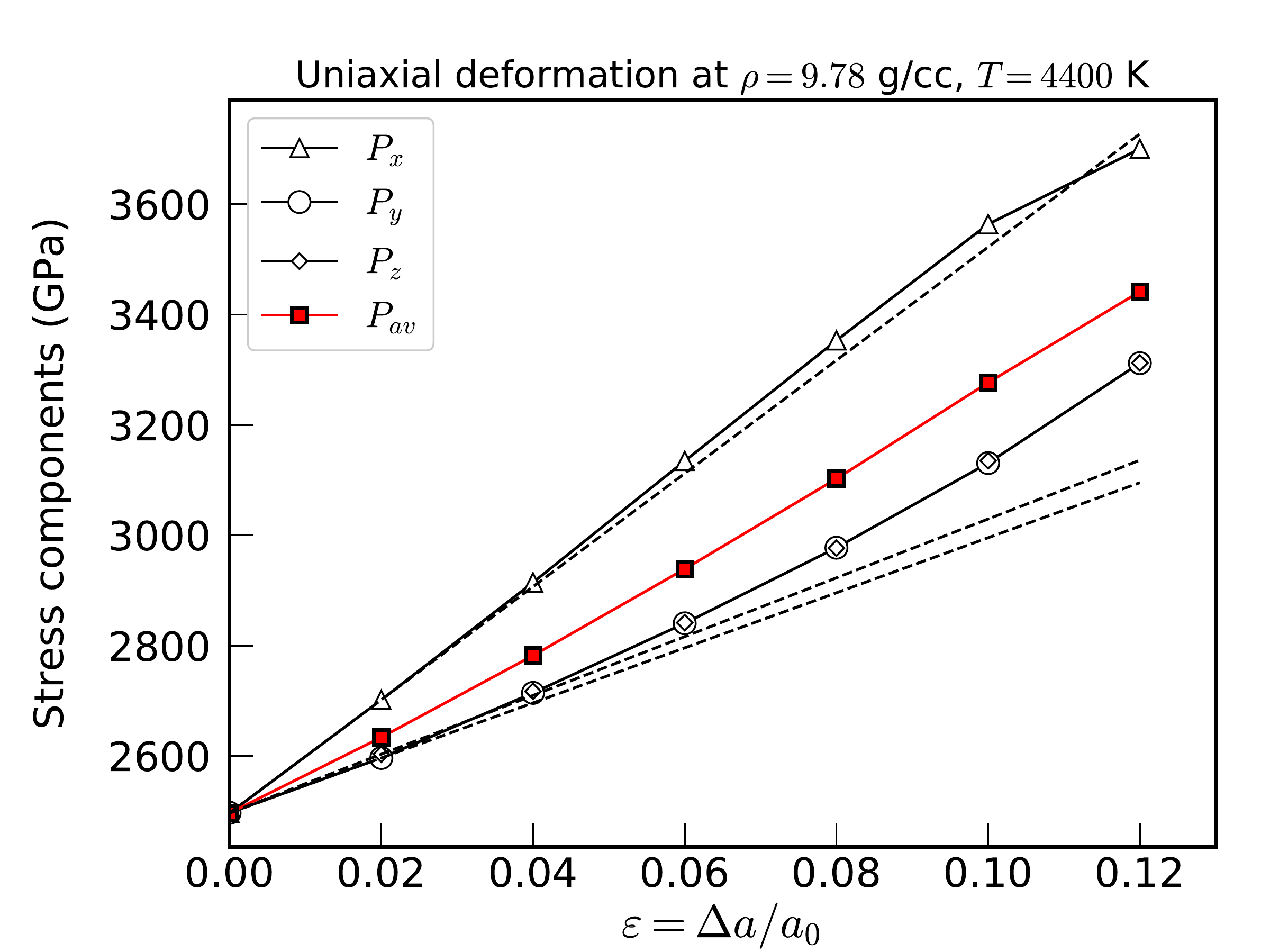}
\caption{Stress-strain curves for diamond (upper panel) and BC8 structures (lower panel). The dashed lines depict the linear regime from where the elastic constants, shown in Fig.~\ref{fig:Cij}, were obtained.}
    \label{fig:CijFit}
\end{figure}

Our simulations of uniaxially compressed cells enable us to determine the elastic constants $C_{11}$ and $C_{12}$ of carbon at extreme $P$-$T$ conditions, as we shown in Fig.~\ref{fig:Cij}. Since diamond and BC8 are cubic structures, one finds $C_{11}=C_{22}=C_{33}$ and $C_{12}=C_{13}$. In Fig.~\ref{fig:Cij}, we show the calculated the elastic constants for both structures, obtained from a set of uniaxial distortions performed at different conditions. We observe that both $C_{11}$ and $C_{21}$ depend strongly on pressure, but there is no significant dependence on temperature. For diamond, the value of $C_{11}$ at 650~GPa is almost three times higher than its value at zero pressure (1161~GPa). A similar behavior is observed in BC8, where the rate of increase of $C_{11}$ with pressure is approximately 2.96, smaller than the value of 3.54 for diamond. Based on a linear fit, we estimate $C_{11}$ to jump from 4883 GPa to 6354 GPa at 1000 GPa, where diamond transforms into BC8. In Fig.~\ref{fig:Cij}, we highlight the region where our computed elastic constants violate the generalized Born stability criterion, that requires the determinant of the stiffness tensor to be positive~\cite{Wang1993}. For cubic crystals, this requires, $C_{11}-C_{12}>P$, which is no longer satisfied for our simulations at the highest pressures in this plot.
In Fig.~\ref{fig:CijFit} we show the stress-strain curve for diamond at ambient conditions and for high pressure conditions in the BC8 phase, to illustrate of how the elastic constant were obtained.
Knowing the elastic constants at high pressure is fundamental to developing strength models~\cite{Orlikowski2008}, and they can be used to account for plastic work during compression~\cite{Bradley2009}, which we will discuss in Sec.~\ref{sec:PlasticWork}.


(v) We also considered the effects of porosity by performing constant-density, uniaxial strained simulations of defect-bearing BC8 cells for the same density of 9.78 \gcc\, as above. Using the prescription in Sec.~\ref{sec:SimDetails}, we generated samples with 7.8\% defect-induced porosity (illustrated in Fig.~\ref{fig:porous}) to match Smith et al.'s sample properties~\cite{Smith2014}. The average normal stress (pressure in the case of zero strain) was found to be 2400 GPa at 4400 K, which is 100 GPa lower than the pressure of the perfect crystal at the same temperature-density conditions. For a given average stress, the defect-laden cells have a smaller volume.
One could have expected that this volume reduction leads to higher stresses for given density, making it easier to match the measured stress values, but our results in Fig.~\ref{fig:porous} suggest otherwise. In the bottom left panel, the maximum stress value of strained, defective cells was in fact lower than those we obtained for the defect-free crystals. We find the defects are mobile at a temperature of 4400 K. Their mobility tends to distribute the stresses more evenly, which reduces the structures ability to maintain large stresses in a particular direction. 

(vi) Finally we report the observation that when the experimental density values are rescaled by a factor of 1.08 (yellow shaded region in Fig.\ref{fig:porous}), the stress-density points came into good agreement with our DFT-MD predictions. The 1.08 factor equals the ratio of the density of defect-free diamond, 3.515~\gcc, and the initial density of the porous diamond in the experiment $3.25$~\gcc. This scaling has been shown to work fairly well in shock compression experiments on porous materials~\cite{Militzer2021,Gonzalez-Cataldo2020,ZhangCH2017,Driver2017,DriverNitrogen2016,Soubiran2019}. If the shock Hugoniot curve for an initial density $\rho_0^A$ is known but the curve for the initial density $\rho_0^B$ is needed, one can approximately scale the densities of the original Hugoniot curve by the ratio of $\rho_0^B/\rho_0^A$. Based on this analogy, one may want to multiply the Smith et al. densities by $f=3.515/3.25=1.08$ or to multiply our DFT-MD densities by $1/f$. While this scaling brings the results from experiments and simulations into good agreement, it leaves two open questions. It is unclear why the DFT-MD results would better describe an experiment that started from the density defect-free diamond, and regardless of what the initial density was in the experiments, it should be possible to directly match the reported density-stress points with DFT-MD simulations at some temperature.

It should also be pointed out that ramp experiments are not as well understood as Hugoniot measurements. Nonhydrostatic effects may play an important role and, though Bradley et al.~\cite{Bradley2009} and Smith et al.'s~\cite{Smith2014} experiments provide measurements of the longitudinal stress as a function of density, they do not report any other stress components. The deviations between different stress components could be as large as 45\% for pressures below 1 TPa, as we will discuss in the next sections.
For the BC8 phase, a reduction of 15\% in the lattice parameter at 2500 GPa can cause the longitudinal stress to be 12\% larger than the average stress, as we illustrate in Fig.~\ref{fig:porous}.
All of these results and open questions motivate us to construct a thermodynamic model to illustrate how heating may be introduced in such experiments.

\section{Plastic Work}\label{sec:PlasticWork}

\subsection{Infinitesimal heating due to plastic work}

It has been suggested that the temperature increase in ramp compression can be explained as a temperature increase due to isentropic compression combined with additional heating from plastic work~\cite{Fowles1961,Bradley2009,Ping2013,Orlikowski2008,Fratanduono2020}. Here, the increment in plastic work done per unit volume in the plastic flow region is given by
\begin{equation}\label{eq:dWp}
\frac{dW_p}{V_0}=\sigma_{ij}'d\varepsilon_{ij}^p =  \bar\sigma d\varepsilon^p,
\end{equation}
where $\sigma_{ij}'=\sigma_{ij}-\sigma_p\delta_{ij}$ is the deviatoric stress of the stress tensor, $\sigma_{ij}$.  The average stress is given by $\sigma_p=(\sigma_{xx}+\sigma_{yy}+\sigma_{zz})/3$. The equivalent stress for plastic deformations, $\bar\sigma\equiv\sqrt{\frac32\sigma_{ij}'\sigma_{ij}'}$~(Einstein summation implied), is often assumed to lie on the yield surface through the von Mises criterion, $\bar\sigma=Y$, where $Y$ is assumed to depend only on volume compression, and is taken from the flow stress reported by Bradley \emph{et al.}~\cite{Bradley2009}. Thus,
\begin{equation}\label{eq:dWp_eta}
\frac{dW_p}{V_0}=Y(\eta)\frac{d\varepsilon^p}{d\eta}d\eta,
\end{equation}
where $\eta\equiv\rho/\rho_0>1$. The total plastic work done is then given by
\begin{equation}\label{eq:Wp_eta}
\Delta W_p=V_0\int_1^\eta Y(\eta')\frac{d\varepsilon^p}{d\eta'}d\eta'.
\end{equation}
The relative plastic strain, $d\varepsilon^p/d\eta$, can be related to the volumetric compression, $\eta$, through~\cite{Ping2013}
\begin{equation}
    \varepsilon^p = \frac{2}{3}\left[\varepsilon_{11}(\eta)-\frac{Y(\eta)}{2G(\eta)}\right] \quad,
\end{equation}
where $\varepsilon_{11}=\ln(\rho/\rho_0)=\ln(\eta)$ and $G$ is the shear modulus that is taken from the Steinberg-Guinan model~\cite{Orlikowski2008}.
This is a simplistic model of plasticity, with all of the plastic work going into heating the system. The corresponding temperature rise due to this irreversible processes is
\begin{equation}\label{eq:Trise}
\Delta T_{\text{rise}}=\int_0^{\Delta W_p} \frac{dE}{C_V}.
\end{equation}
The final temperature reached in this model of ramp compression is given by the contribution from isentropic compression, $\Delta T_{\text{isen}}$, and the contribution from plastic work,
\begin{equation}\label{eq:Tfinal}
T_{\text{final}}= \Delta T_{\text{isen}}+\Delta T_{\text{rise}}.
\end{equation}
The Gr\"uneisen parameter gives the temperature increase under isentropic compression, with $\gamma\rho=\gamma_0\rho_0$ assumed constant, which  results in $\Delta T_{\text{isen}}=T_0\exp\left(\gamma_0\left(1-\rho_0/\rho\right)\right)$, with $\gamma_0$ the ambient-density Gr\"uneisen parameter.

\begin{figure}[!h]
\includegraphics[width=8cm]{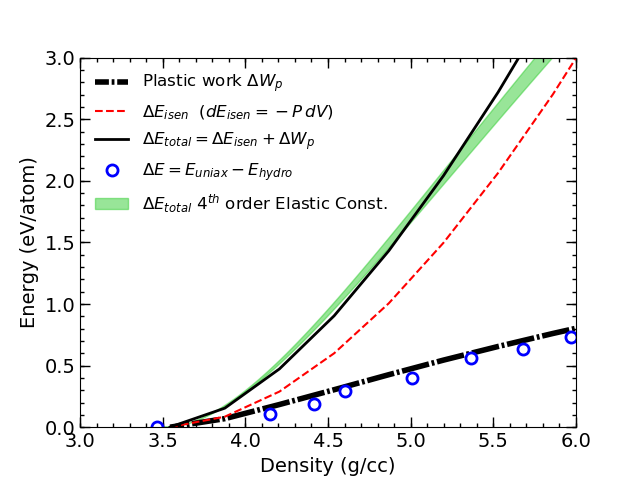}
\caption{Plastic work (dot dashed) and energy increase upon isentropic compression (solid black line), compared to the energy difference between hydrostatically and uniaxially compressed diamond (open blue circles). The energy difference with the isentrope, $\Delta E_{\rm isen}=-\int P\, dV$ (dashed red line) is added to the plastic work to account for the total energy increase $\Delta E_{\rm total}$. Estimates from elasticity theory using 4th order elastic constants from Telicho \emph{et al}.~\cite{Telichko2017}. }
\label{fig:Wp}
\end{figure}

In Fig.~\ref{fig:Wp}, we compare the plastic work done to compress diamond to a given density with the DFT-derived energy difference, $\Delta E$, between uniaxially and hydrostatically compressed diamond to the same density from ambient conditions.
We find that the energy excess, $\Delta E$, is comparable to our plastic work estimates, obtained from Eq.~\eqref{eq:Wp_eta}. When one adds together the energy of isentropic compression and $\Delta E$, one obtains the solid black curve in  Fig.~\ref{fig:Wp} that represents the total energy increase in this ramp compression model.

Alternatively, we can use elasticity theory to estimate energy increase due to uniaxial compression. From the elastic constants of diamond reported by Telichko \emph{et al.}~\cite{Telichko2017}, we calculated this energy considering the error bars in elastic constants,  shown by the shaded green curve in Fig.~\ref{fig:Wp}. We observe that there is good agreement between the total energy increase, $\Delta E_{\text{total}}$, obtained from the plastic work model and that from elasticity theory.

\section{Ramp Compression Model}\label{sec:RWCmodel}

Our DFT-MD simulation results motivated the development of a model that can explain the rise of temperature during ramp compression. In some experiments~\cite{Bradley2009,Smith2014,Smith2018}, the main ramp pulse may be preceded by an initial shock, which takes the sample to a point along the principal shock Hugoniot curve, from where it is ramp compressed. Here, we model the energy transfer that occurs in solid materials as they are compressed by a ramp wave.

In ramp compression, a ramp pulse creates a uniaxial strain in the material, that eventually relaxes to a hydrostatically compressed state at conditions exceeding the dynamic yield stress~\cite{Swift2008}. This relaxation increases the internal energy and entropy. It is a complex process that involves length and time scales that are orders of magnitude larger than the scope of existing ab initio simulation methods. Furthermore, since strain is uniaxial, only the longitudinal component of the stress can be measured~\cite{Bradley2009,Smith2014}, and access to the full stress tensor and thus also the average stress 
is only available through models~\cite{Smith2018}. For this reason, we need to design an analogue process that captures the essential stages of ramp compression, while being predictive and amendable to ab initio computations.

\begin{figure}[!hbt]
    \includegraphics[width=8cm]{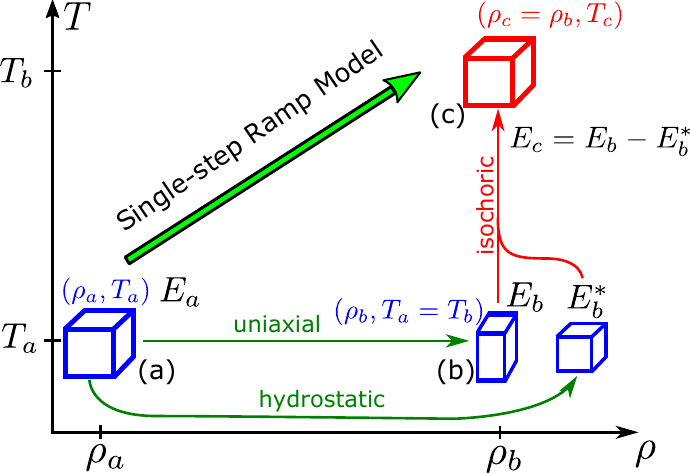}
    \includegraphics[width=6cm]{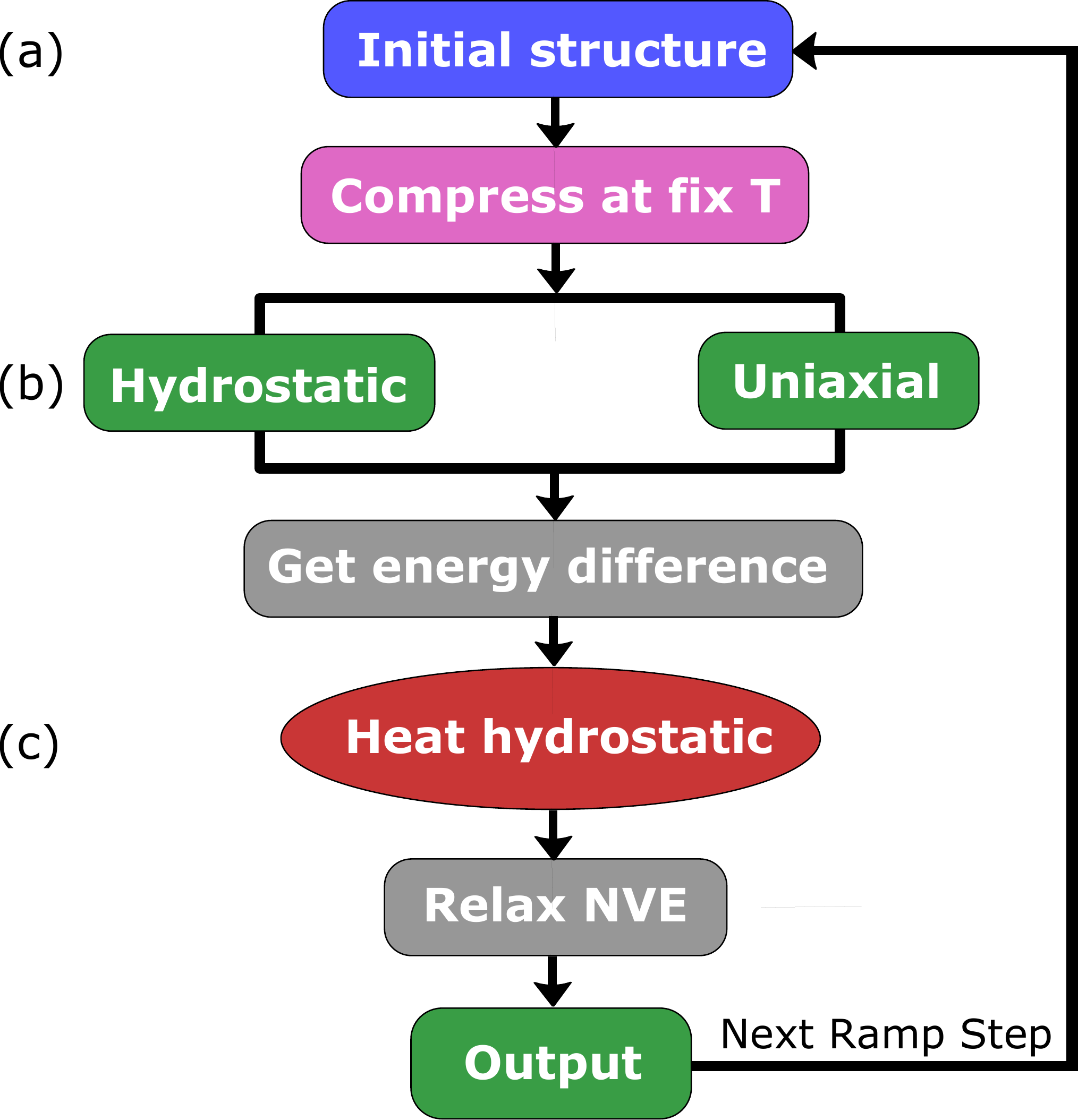}
    \caption{Illustration of a single step of our ramp wave compression model. The system is uniaxially compressed from the hydrostatic state (a) to the uniaxially compressed state (b) along an isotherm.
    The energy of this state, $E_b$, which is higher than that of the equivalent hydrostatic state at the same temperature and density, $E_b^*$, are used to obtain the energy difference, $E_c$, which is given to the hydrostatically compressed sample at this density as kinetic energy.} 
    \label{fig:scheme1step}
\end{figure}

In our model, depicted in Fig.~\ref{fig:scheme1step}, we approximate ramp loading as a series of compression and relaxation steps.
In each step, we perform DFT-MD simulations in the NVT ensemble to prepare a cubic sample that is (a) prepared under hydrostatic conditions at a given density and temperature and then (b) uniaxially compressed to a higher density. The internal energy is compared to that of a cubic sample under hydrostatic conditions at the same density and temperature. We derive the internal energy difference between the two states, which we found to be equivalent to the plastic work (Fig.~\ref{fig:Wp}), in accord with our assumption that relaxation from the uniaxially compressed state to hydrostatic conditions is irreversible and therefore causes heating during ramp-compression. Thus, (c) we add the energy difference to the energy of the hydrostatically compressed state and let the system equilibrate to a new temperature while holding the energy constant. To achieve this, we performed DFT-MD simulations in the microcanonical (NVE) ensemble. Step (c) represents the relaxation to the hydrostatic state, with the elastic energy accumulated in the metastable, uniaxially compressed state being transformed into internal energy, increasing the sample temperature as hydrostatic conditions are achieved.

Since the energy is held constant after adding this energy difference, the system reaches a new thermodynamic equilibrium state with a higher temperature. This new state is taken as the initial state for the next compression step, and the entire process (a)-(c) is repeated. Every compression step consists of these three stages. The total number of compression steps is chosen so that they lead to a reasonable subdivision of the pressure interval spanning the initial and final compression.

\begin{figure}[!hbt]
    \includegraphics[width=8cm]{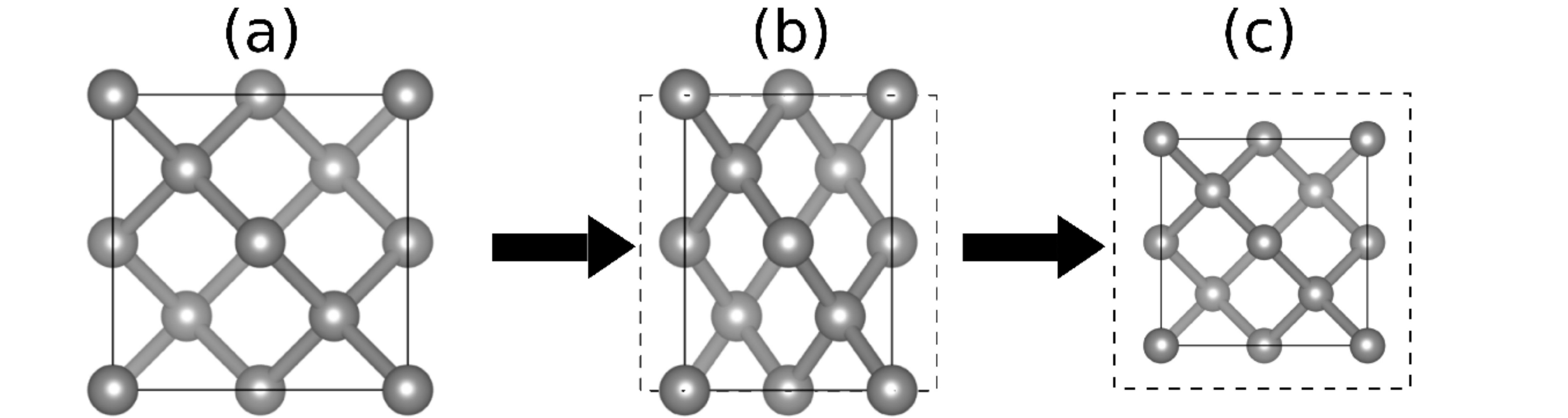}
    \includegraphics[width=8cm]{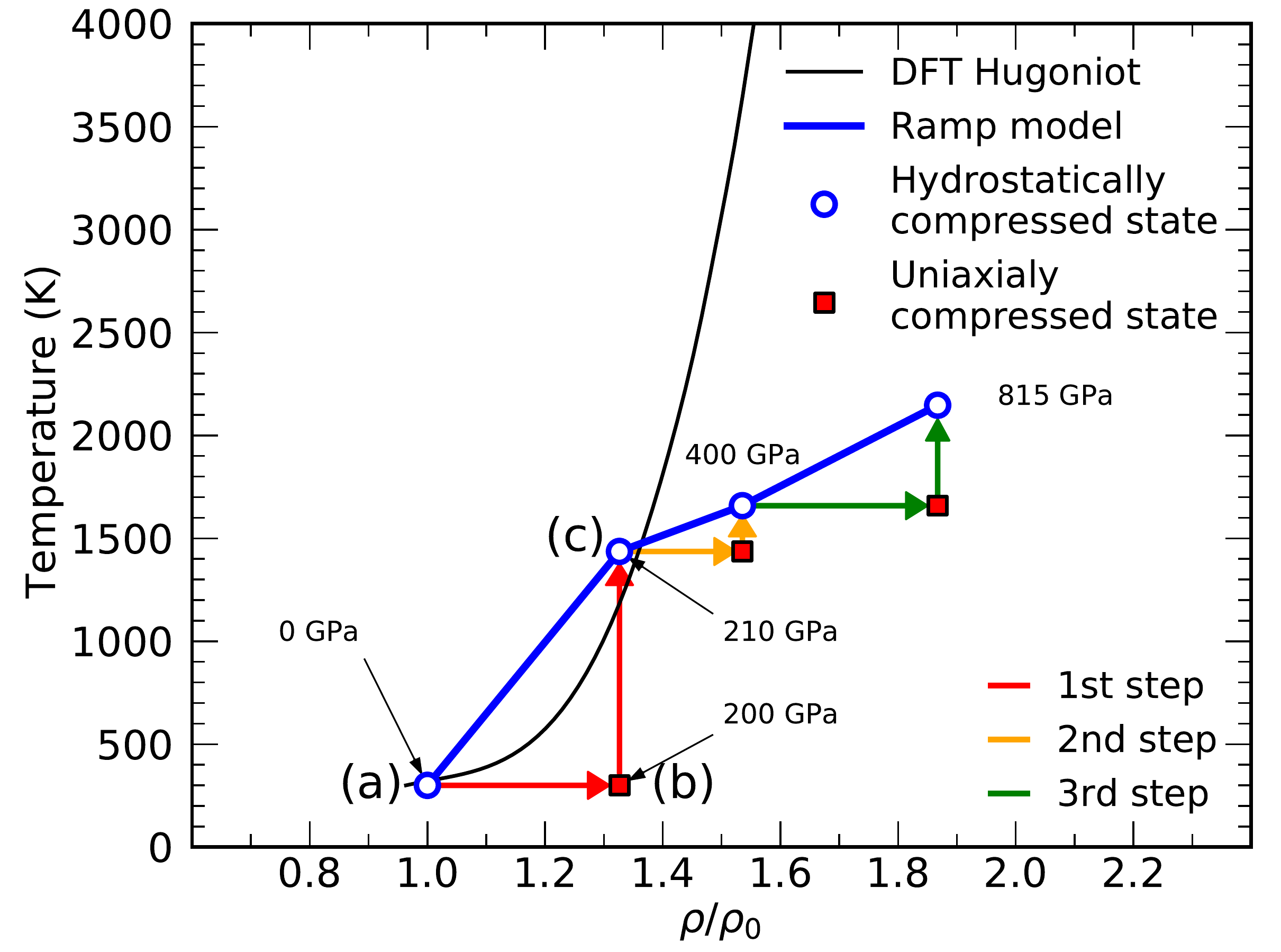}
    \caption{Illustration of our ramp wave compression model for diamond, using three compression steps. The system is uniaxially compressed to the next density (red squares) along an isotherm (horizontal arrows), and then the energy difference with respect to the equivalent hydrostatically compressed state (at the same conditions), is given to the cubic sample as kinetic energy (vertical arrow). The relaxation of the system after increasing energy is performed in the microcanonical ensemble (NVE), which leads to the new equilibrium temperature (blue circles). The cycle is repeated from this point using the equilibrated state as the new initial state for the next step. A second and then a third compression step is performed in order to reach 800 GPa.}
    \label{fig:scheme3steps}
\end{figure}

We show an application of our model to diamond using the three consecutive compression steps that we illustrate in Fig.~\ref{fig:scheme3steps}. In the first step, the sample is  (a) prepared at 300 K and ambient density of $\rho_0\equiv 3.468$ \gcc\, and taken to (b), the uniaxially compressed state, at the same temperature. The corresponding energy difference at these conditions is given to the sample hydrostatically compressed to the same density and temperature, a process that takes the system to (c). A second step is initiated from this state to reach a compression of 1.5-fold, and then a third step takes the diamond to 1.86-fold compression, which is the density needed in order to reach ~800 GPa. In this case, we have chosen the densities in each step to closely match 200, 400, and 800 GPa. The resulting curve, shown in open blue circles in Fig.~\ref{fig:scheme3steps}, is the predicted ramp compression path in temperature-density space.

\subsection{Single-step calculations}

\begin{figure}[!hbt]
    \includegraphics[width=8cm]{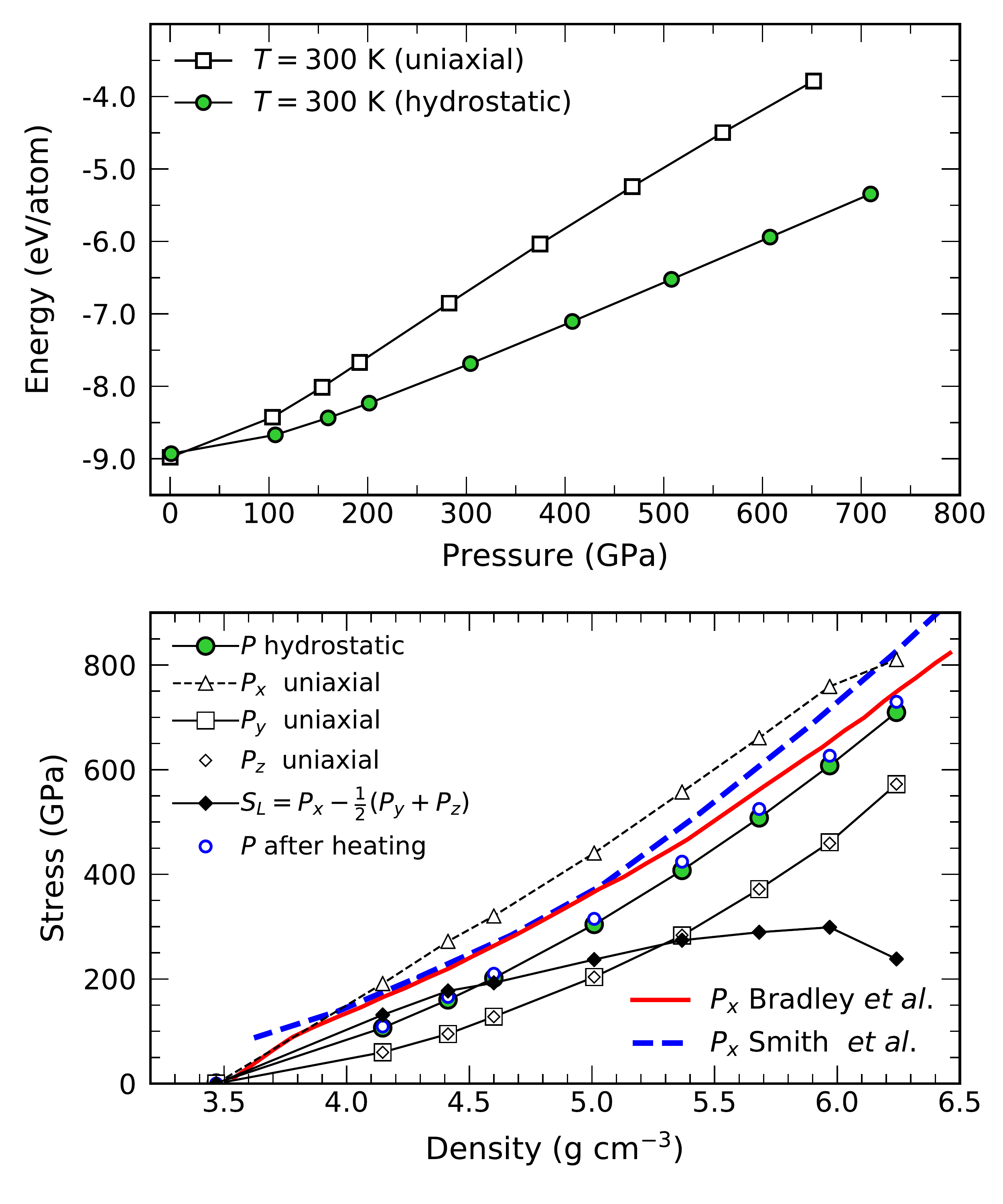}
    \caption{Total energy and stress in hydrostatically (solid green circles) and uniaxially compressed (open squares) diamond along the 300 K isotherm for a series of single-step uniaxial compressions. Ramp compression experiments by Bradley et al.~\cite{Bradley2009} and Smith et al.~\cite{Smith2014} are shown for comparison. The cubic, hydrostatically compressed state is energetically favored over the uniaxially compressed state for all pressures. Along this isotherm, the pressure in hydrostatically compressed samples, $P$, is always below the experimental curves. The open blue circles represent the heated sample (stage (c) in our ramp model) for each of the individual single-step compressions.  The drop in stress anisotropy, $S_L=P_x-1/2(P_y+P_z)$, at $\rho/\rho_0=1.8$ indicates the onset of a mechanical instability in the uniaxially compressed crystal that is enhanced by thermal motion of the nuclei.}
    \label{fig:EnergyDensity}
\end{figure}

\begin{figure}[!hbt]
    \includegraphics[width=8cm]{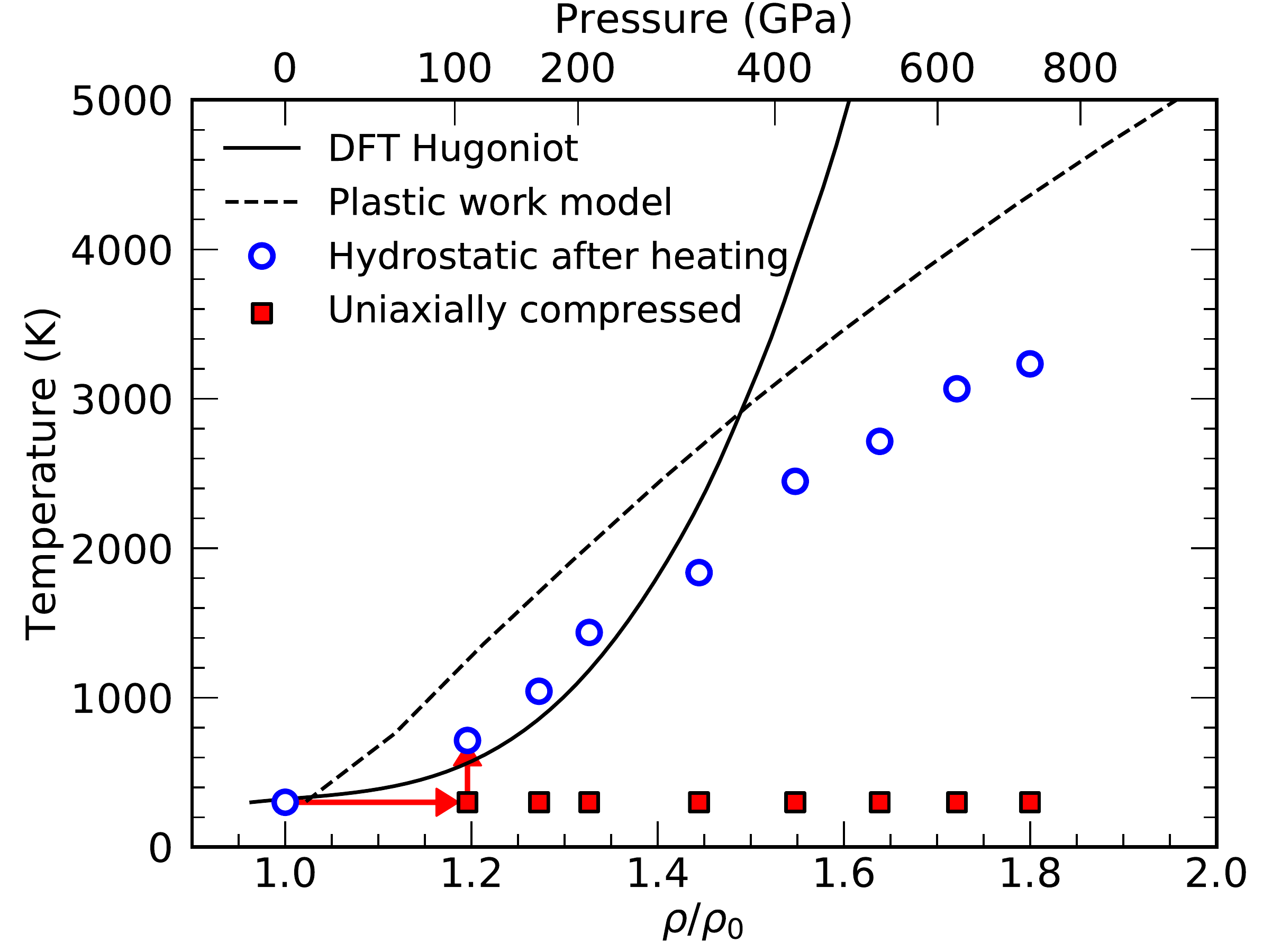}
    \caption{Final temperature (relaxation to hydrostatically compressed stage (c)) vs compression ratio for a series of single-step uniaxial compression steps, corresponding to the blue circles in Fig.~\ref{fig:EnergyDensity}. The final temperature from our single-step ramp model is compared the temperature rise derived from the plastic work model using Eq.~\eqref{eq:Tfinal}.}
    \label{fig:SeriesOf1steps}
\end{figure}

In order to determine the dependence on the step size in our model (going from (a) to (b)), we subjected diamond to a series of single-step uniaxial compressions from initial ambient conditions, as shown in Fig.~\ref{fig:EnergyDensity} and ~\ref{fig:SeriesOf1steps}. In Fig.~\ref{fig:EnergyDensity}, we compare the total energy and pressure of the uniaxially and hydrostatically compressed cells along the 300 K isotherm. Each point corresponds to a different uniaxial compression ratio (and thus, to a different of target density) in going from (a) to (b) (see Fig.~\ref{fig:EnergyDensity}). The uniaxial compression is achieved by reducing the lattice constant in the [100] direction, while keeping the other cell vectors unchanged. We compare our results with two reports of $P_x$ as a function of density during ramp compression (Bradley et al.~\cite{Bradley2009} and Smith et al.~\cite{Smith2014}). The change in slope of Bradley et al.'s curve near $\rho=3.8$ \gcc\, (1.1 $\rho_0$) indicates the elastic response limit.

 As expected, the energy of the hydrostatically compressed cell is always lower than that of the cell subjected to uniaxial compression, which reflects the fact that the latter is a state of metastable equilibrium. The hydrostatic pressure in the cubic cell, however, is higher than the average stress, $P_{av}=(P_x+P_y+P_z)/3$, in the uniaxially compressed cell, along this isotherm. The stress along the compression direction, $P_x$, is about 35\% higher than the average stress,  $P_{av}$, and 65\% higher than the perpendicular stress, $\frac12(P_y+P_z)\approx P_y\approx P_z$, as shown in Fig.~\ref{fig:EnergyDensity}. When the density reaches 6.24 \gcc\, (1.8 $\rho_0$) in hydrostatic compressions at 300 K, the lattice constant is 18\% smaller than its ambient value of 3.57\,\AA, while this requires a compression of 45\% of the strained axis of the uniaxially compressed cell.
 Beyond this point, further compression is no longer feasible, since a longitudinal stress instability triggers a crystal structure collapse in this highly strained cell. The decrease in the slope of $P_x$ at $1.8 \rho_0$ and subsequent decrease in the longitudinal stress anisotropy, $S_L=P_x-\frac12(P_y+P_z)$, is a signature of the instability growing in the uniaxially distorted crystal.

After adding the energy excess of the strained cells, shown in Fig.~\ref{fig:EnergyDensity}, to each of the corresponding cubic cells (stage (c) of our ramp compression model), we can calculate how much the temperature increases for each of the single-steps compressions. The resulting temperatures are shown in Fig.~\ref{fig:SeriesOf1steps}, where the shock Hugoniot curve obtained from DFT simulations~\cite{Correa2008} is also shown for comparison. Beyond 1.4-fold compression (~300 GPa), the final temperatures under ramp loading start deviating from shock Hugoniot curve in temperature-density space. This can be interpreted as an upper limit for single-step uniaxial compressions in our ramp model, since we expect the first step to take the system to a point close to the Hugoniot curve, because the energy used to compress the sample in a shock is then released to heat the sample. 
The resulting temperatures are lower that those obtained from the plastic work model presented in Sec.~\ref{sec:PlasticWork}.

\subsection{Multi-step calculations}

\begin{figure}[!hbt]
    \includegraphics[width=8cm]{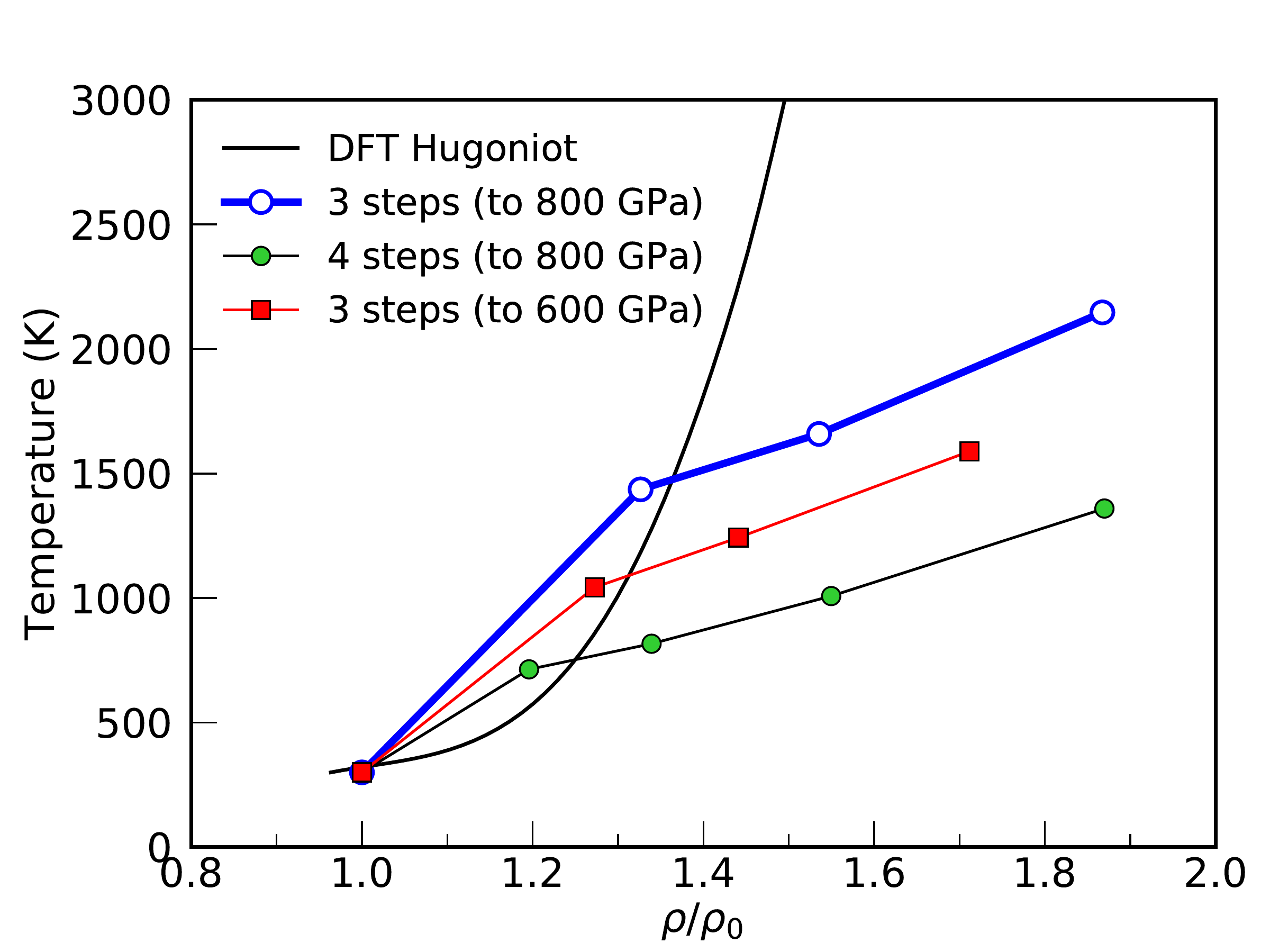}
    \caption{Temperature rise in our ramp-compression model for different compression paths. The final temperature depends on the number of steps taken to subdivide the pressure interval. For a final density of 6.48~\gcc\,(800 GPa), a 3-steps subdivision leads to a final temperature of 2000 K, while 4-steps lead to 1200 K. For a final pressure of 600 GPa, a 3-steps sequence results in a final temperature of 1500~K. }
    \label{fig:Trise}
\end{figure}

After determining the temperatures of a series of independent single-step compressions, we applied several consecutive compression steps to diamond, starting from ambient conditions and studied how our predicted final temperatures depend on the number of steps chosen. We first subdivided the pressure interval 0 to 800 GPa in 3 steps: 0 GPa$\to$200 GPa$\to$400 GPa$\to$800 GPa, as we show in Fig.~\ref{fig:Trise}. For this sequence, we obtain a final temperature of 2150 K at 800 GPa. As the number of steps is increased over a given pressure interval, one expects to generate less heat because smaller uniaxial compression is applied at every step and, consequently, less energy is available to increase the temperature. Indeed, this is what we observe in Fig.~\ref{fig:Trise} when we use four compression steps: 0 GPa$\to$100 GPa$\to$200 GPa$\to$400 GPa$\to$800 GPa. Using four steps, then, leads to a lower temperature of 1400 K. For a very large number of compression steps, this curve approaches an isentrope, as the amount of heating decreases for smaller steps. We also show a different 3-steps sequence for a final pressure of 600 GPa, considering 150 GPa and 300 GPa as the intermediate steps. At the final pressure of 600 GPa, we obtain a temperature of 1500 K.

Therefore, at 800 GPa, our model predicts a temperature of 3300 K when we use a single step (Fig.~\ref{fig:SeriesOf1steps}), 2150 K for 3 steps, and 1400 K for 4 steps. Bradley et al. estimated that this temperature should be 6300 K at this pressure~\cite{Bradley2009}, assuming that heating responds to an isentropic plastic flow process. While all these temperature estimates are based on different assumptions, they all lie below the melting curve and support the conclusion that the sample remained solid at these conditions~\cite{Bradley2009}. The appropriate number of steps in our model can be determined once temperature measurements are provided in a ramp-compression experiment, which will also provide information about the strain in the material.

\section{Multi-shock Hugoniot calculations}\label{sec:Multishock}

In order to provide a benchmark for our model calculations, we performed a multi-shock Hugoniot analysis over a dense grid of DFT-MD data in the 0$-$5000 GPa pressure range, considering also the high-pressure BC8 and simple cubic (SC) phases of carbon, and then compare our results with the values reported by Smith et al.~\cite{Smith2014}.  A multi-shock Hugoniot offers one approximation to a ramp wave, which can be thought a succession of shocks. By breaking up a single shock into multiple smaller shocks, one approaches an isentrope~\cite{Driver2018}, and we expect the conditions in each of these shocks to be similar to those obtained by our ramp model. When a shock passes through a material, the initial energy, pressure, and volume $(E_0,P_0,V_0)$ and the final values $(E,P,V)$ are related by the Rankine-Hugoniot equations~\cite{Benedetti1999},
\begin{equation}\label{eq:Hugoniot}
    (E-E_0)=\frac12(P+P_0)(V-V_0)
\end{equation}
which describes the conservation of mass, momentum, and energy across the shock front. In the multi-shock scheme, a point on the principal Hugoniot curve is used as the initial state for a second shock. A point on this secondary Hugoniot curve is chosen as an initial state for the third shock, and so on. Since Eq.~\eqref{eq:Hugoniot} needs to be evaluated at many volumes $V$ for a given temperature $T$ to obtain $E(V,T)$ and $P(V,T)$, a dense volume-temperature grid is required to find the states that satisfy this equation.

\begin{figure}[!hbt]
    \includegraphics[width=8cm]{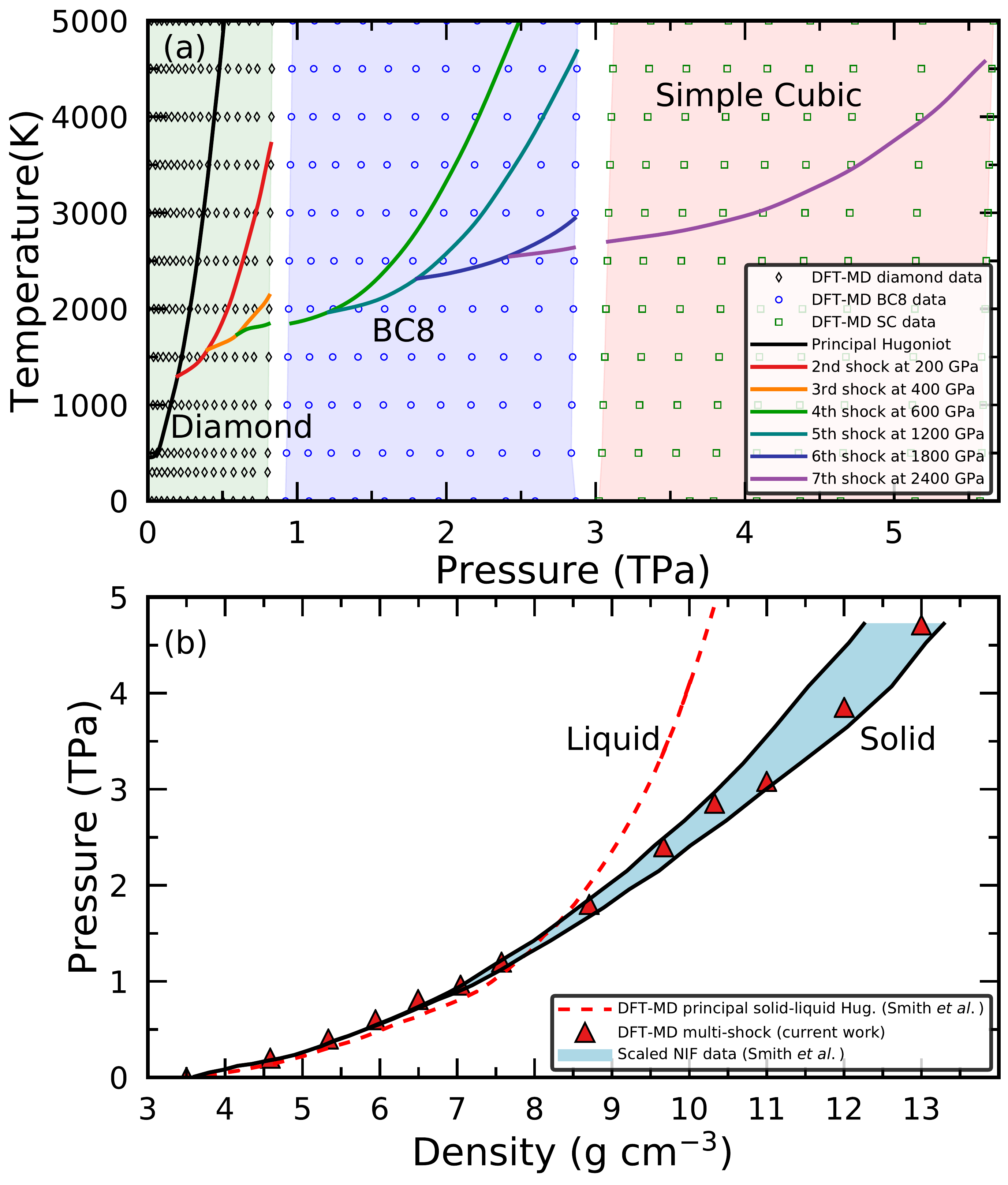}
    \caption{Multiple-shock Hugoniot curves of carbon at pressure-temperature conditions of the diamond, BC8, and simple cubic phases. The shocks occur in diamond at 200, 400, and 600 GPa, and in BC8 at 1200, 1800, and 2400 GPa. In the bottom figure, the multi-shock points of the panel above are plotted in pressure-density space, and compared with the experimental ramp-compression values of Smith et al.~\cite{Smith2014} and the DFT-MD solid-liquid principal Hugoniot of diamond~\cite{Benedetti1999,Correa2008}. The Smith et al. densities were scaled by a factor of 1.08. 
    }
    \label{fig:MultipleShocks}
\end{figure}

In Fig.~\ref{fig:MultipleShocks}, we show the full temperature-pressure range at which we  performed DFT-MD multi-shock calculations. The isotherms span across the diamond, BC8 and SC phases of carbon. Seven consecutive shock Hugoniot curves were generated as described above. The initial state for each curve was chosen such that the pressures matched closely 0, 200, 400, 600, 1200, 1800, and 2400 GPa. The initial conditions in each shock form a curve that is shallower than the principal Hugoniot curve, as expected. These are shown in the pressure-density diagram of Fig.~\ref{fig:MultipleShocks}, which shows that they closely match the data from Smith et al.~\cite{Smith2014}.

\subsection{Ramp Model versus Multi-shock Hugoniot curves}

We applied our ramp-compression model to the 0$-$5000 GPa pressure interval, choosing an evenly distributed set of compression steps: 0, 300.25, 612.15, 1225, 1850, 2450, and 4000 GPa. We added an intermediate step at 1000 GPa, where diamond is predicted to transform to the BC8 structure~\cite{Benedetti1999,Benedict2014,Martinez2012}. For the additional step at 1000 GPa, we use the density and temperature reached in the diamond structure to prepare the BC8 structure, and then we uniaxially compressed BC8 to 1225 GPa. This prevents a spontaneous phase transition during the uniaxial compression. However, for the additional step at 3000 GPa, it was not possible to stabilize the simple cubic crystal at the corresponding temperature under uniaxial compression, which can be an indication of a BC8 to simple hexagonal transition~\cite{Martinez2012}. Therefore, the last step to 4000 GPa was also performed in the BC8 structure.

\begin{figure}[!hbt]
    \includegraphics[width=8cm]{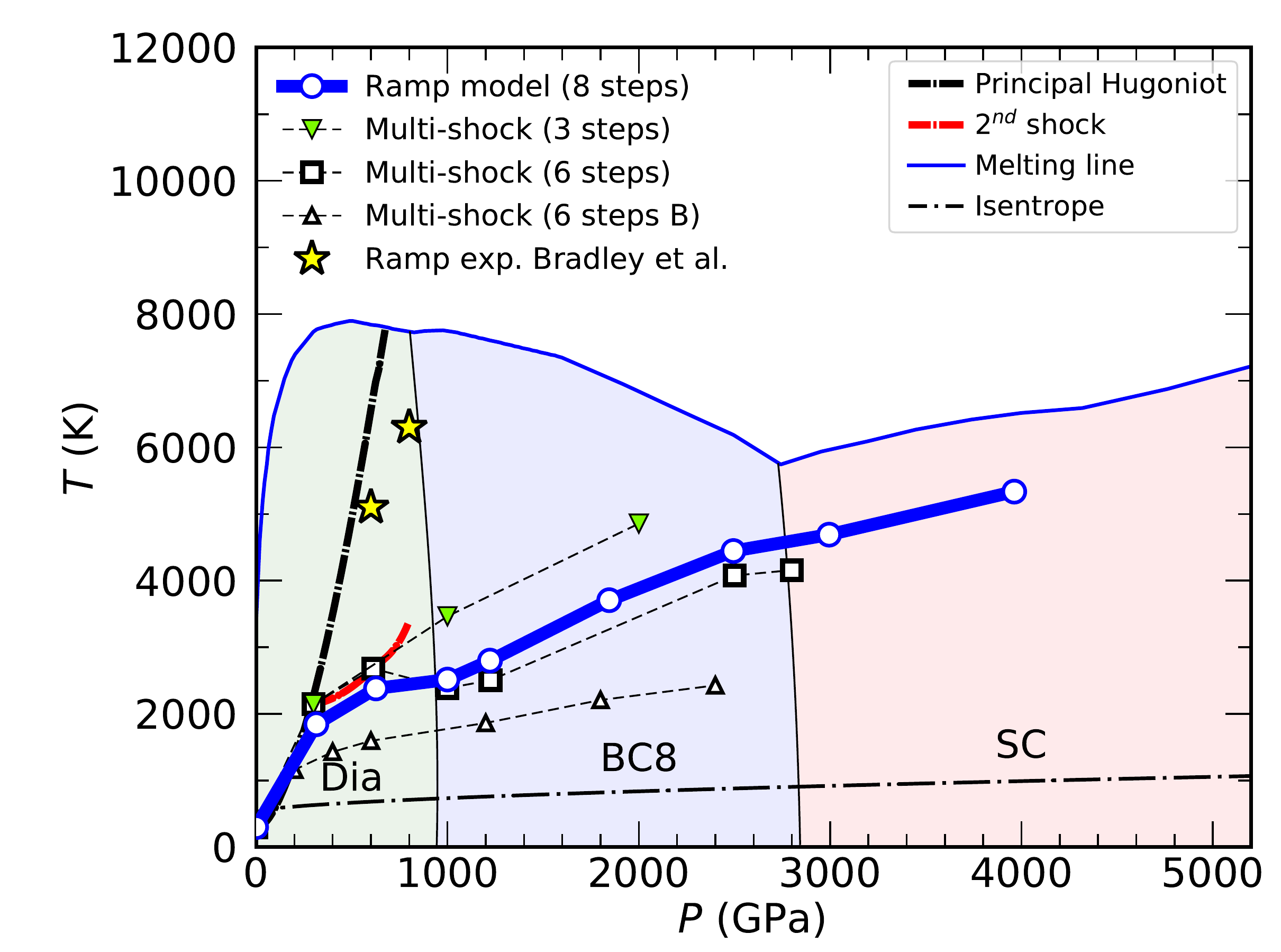}
    \caption{Temperatures predicted by the multi-shock scheme compared to our ramp model in the 0$-$5000 GPa pressure range, using 8 steps. Intermediate steps are performed at 1000 and 3000 GPa, where solid/solid phase transitions are predicted to occur. All uniaxial compressions from 1000 to 4000 GPa were performed in the BC8 structure. The melting curve (thin blue curve)~\cite{Benedict2014} is shown for comparison. The dot-dashed black curve represents the isentrope that crosses the principal Hugoniot curve at 110 GPa, the diamond elastic limit. Temperature estimates, based on plastic work, at 600 and 800 GPa from the ramp compression experiments of Bradley \emph{et al.}~\cite{Bradley2009} are shown as yellow stars.}
    \label{fig:MultipleShocksvsRamp}
\end{figure}

We compared the results from our ramp model using 8 steps, with the multi-shock scheme shown in Fig.~\ref{fig:MultipleShocksvsRamp}, and steps chosen to closely match the pressure sequence of the ramp-loading model. We observe that the final temperatures predicted by the two models are in good agreement, and below the melting curve. In our ramp model, we find a temperature of 4600~K when the pressure reaches 3000~GPa, while this temperature is 4200~K for this same pressure in the multi-shock approach. The good agreement between the two approaches shows that, in fact, our ramp compression model is equivalent to a series of consecutive shocks. Moreover, our results agree with values from a recent EOS model for diamond~\cite{Swift2020}, based on flow stress estimates and plastic work done in the sample (Fig.~\ref{fig:3Multipleshock}).

Our findings show that, as the number of compression steps increases, the temperature decreases. Most of the heat is delivered in the first shock, after which the temperature increases moderately, when a sufficiently large number of steps is considered. The total number of steps required to predict the temperature under ramp loading depends especially on the magnitude of the first step along the principal Hugoniot curve, which is set by experiments~\cite{Smith2014}.

\subsection{Multi-shocks in the liquid regime}

\begin{figure}[!hbt]
    \includegraphics[width=8cm]{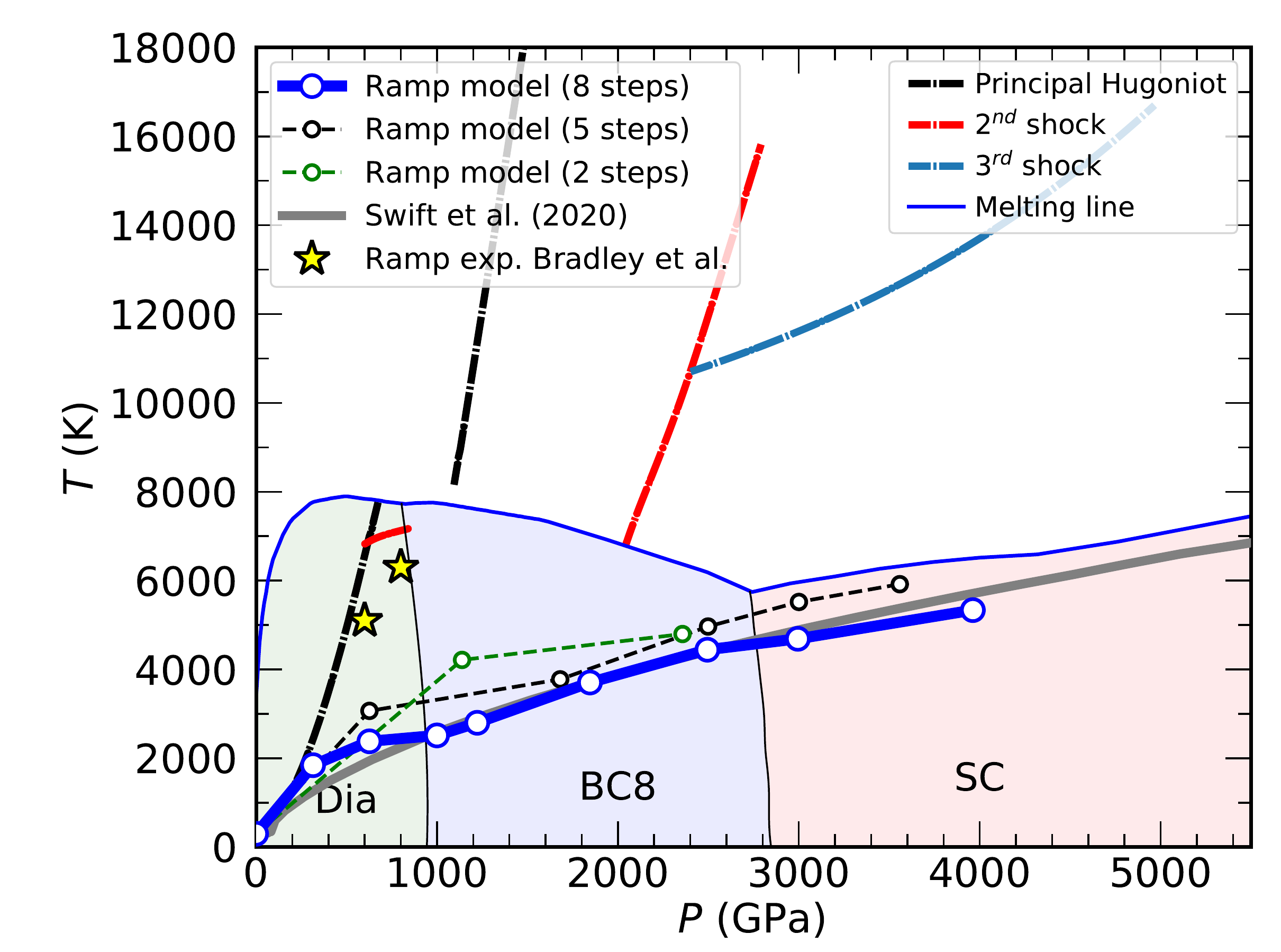}
    \caption{Three-shock sequence in the multi-shock scheme, launching a secondary shock at 600 GPa. The secondary shock continues in the liquid phase at 2000 GPa, while the third shock, that starts at 2450 GPa, reaches 14000 K at 4000 GPa. The second shock crosses the  melting line near 900 GPa. Different combinations of steps in our ramp compression model show that it is not possible to melt the sample with any combination of steps. Temperature estimates, based on plastic work, at 600 and 800 GPa from the ramp compression experiments of Bradley \emph{et al.}~\cite{Bradley2009} are shown as yellow stars. Temperature estimates from Swift \emph{et al}.~\cite{Swift2020} are shown in solid thick grey line.}
    \label{fig:3Multipleshock}
\end{figure}

In Fig.~\ref{fig:3Multipleshock}, we show how different the compression path is when the second shock is launched at 600 GPa, using the multi-shock scheme. First, we observe a discontinuity in the Hugoniot temperature upon crossing the melting line, caused by the heat of fusion. This discontinuity is also observed in the secondary shock, which starts in the diamond solid phase at 600 GPa, which crosses the melting line around 900 GPa and continues in the liquid regime around 2000 GPa. If a third shock is launched from this secondary Hugoniot curve in the liquid phase at 2450 GPa, the temperature can reach values as high as 14000 K at 4000 GPa. Therefore, even if we consider more steps, the final temperature is higher than the melting temperature for this particular choice of the first step. Once a liquid state has been reached, the remaining compression may be close to isentropic (see Fig.~\ref{fig:PhaseDiagram}).

\section{Conclusions}\label{sec:Conclusions}

In order to match the density-stress measurements of Smith et al., we performed DFT-MD simulations of isotropically and uniaxially compressed solid carbon at various temperatures. Simulations of solid structures with vacancies and for the liquid phase were also performed. 
We find the inferred ramp compression path to have a pressure-temperature slope similar to that of the isentropes. Our simulations of solid diamond show that the stresses are always lower when compared with the experimental results for the same density. The experimental stress-density data points can be matched with simulations of liquid carbon but this implies unexpectedly high temperatures, and no signature of melting are reported by Bradley~\emph{et al.} or Smith \emph{et al.}. Still, based on our simulations, we estimate that 30\% of the deposited shock energy was converted into heat while 70\% went into compressing the sample, which means one cannot rule out a priori that the sample has melted. On the other hand, the recent x-ray diffraction measurements by Lazicki et al. demonstrated that the sample did not melt in those experiments.


The question of how hot diamond becomes during ramp compression can in principle be addressed with additional temperature measurements.
However, one can only measure the thermal emission from the sample's free surface, which makes inferring a bulk temperature more uncertain. E.g. Lazicki \emph{et al.} were unable to discriminate between a ramp compression path with a Taylor-Quinney factor (the fraction of plastic work converted to heat) of 0.9 from one with a factor of 0.5, but they suggested that the latter value was more plausible. 


The fact that the reported experimental stresses are higher than the corresponding pressure of isotropically compressed carbon, either diamond or BC8, at any given density is a consequence of one or both of the following: i) the samples have initial porosity, so have a heterogeneous, high-temperature response; ii) the final stress state is not isotropic, but has a residual uniaxial component. Thermal pressure due to irreversible heating under ramp loading is otherwise too small to explain the differences between experimental data and the results from our DFT simulations. The possibility that the evolution of lattice defects and porosity play an important role, or that the samples had not reached thermodynamic equilibrium during the few nanoseconds that these experiments last, cannot be ruled out. Diamond can withstand substantial uniaxial compression, and the stress in the compression direction can be more than 200 GPa  (up to 3 times) larger than the other two directions according to our simulations,  which means that the average stress reached the experiments may be 30\% smaller than the stress that was reported for direction of compression.


The lowest possible temperature path that a ramp compression can follow is an isentrope, which assumes all changes are thermodynamically reversible. While this assumption may be realistic for liquids, during the compression of crystalline materials some irreversible changes are unavoidable. For example, during phase changes,
the atoms need to re-arrange to form a different
crystal structure~\cite{Williams2020,Godwal2015}. But, even without a phase change, the atoms need to re-arrange once the strain is sufficiently large to exceed the elastic limit under dynamic loading. In both cases, ramp and shock, defects are generated to facilitate the relaxation to a new hydrostatic state and we developed a multi-step model in order to characterize these effects for ramp-compression of solids. We find that single-step compressions in our ramp model yields results in fair agreement with the diamond shock Hugoniot below 300 GPa, which  can be interpreted as plastic deformation taking place in uniaxially compressed diamond. The temperatures predicted by our ramp model are in good agreement with the multiple-shock scheme for an equivalent set of steps, providing some validation of the model. We find a strong dependence of the predicted ramp compression path on the initial step. Future experimental techniques must be developed in order to obtain reliable measurements of temperature during ramp compression, as current techniques such as EXAFS measurements of temperature in Fe, have very large error bars~\cite{Ping2013}. In absence of a fundamental microscopic theory for temperature increase during ramp compression, our model provides a reasonable first estimate. We have found that it is possible to estimate the temperature for experiments by calculating temperatures based either on our ramp model outlined before, or on a multi-shock scheme.

Our study is relevant for characterizing the interiors of giant planets because carbon is a notable constituent of ice giants like Uranus and Neptune and has been proposed to occur in the interiors of exoplanets like 55 Cancri e~\cite{Wilson2014,Madhusudhan2012}.  For certain proto-planetary disks with high carbon content, the chemistry of planetary formation may lead to the formation of carbon-rich planets dominated by carbides and diamond. The properties of carbon at extreme conditions are also needed to construct models for white dwarf stars. Diamond’s singular properties (e.g., high bulk modulus, thermal conductivity, transparency across a broad spectrum of electromagnetic wavelengths) is exploited in static high-pressure experiments and as a means of increasing the initial density of (pre-compressed) samples for dynamic compression experiments.  It exhibits a nearly constant melting temperature at 9-20 Mbar~\cite{Benedict2014} and an exceptionally large dynamic strength~\cite{McWilliams2010}. Furthermore, because of its overall technical and scientific importance, including potential use in laser compression experiments as an ablator and target material, diamond makes it one of most important elements across multiple disciplines.

\begin{acknowledgements}
We thank D. Orlikowski and R. Rudd for providing details about the plastic work model. F.G-C thanks Nicolas Verschueren for his help with some figures.
KPD states that their work was sponsored by an agency of the United States government. Neither the United States government nor any agency thereof, nor any of their employees, makes any warranty, express or implied, or assumes any legal liability or responsibility for the accuracy, completeness, or usefulness of any information, apparatus, product, or process disclosed, or represents that its use would not infringe privately owned rights. Reference herein to any specific commercial product, process, or service by trade name, trademark, manufacturer, or otherwise does not necessarily constitute or imply its endorsement, recommendation, or favoring by the U.S. Government or any agency thereof. The views and opinions of authors expressed herein do not necessarily state or reflect those of the U.S. Government or any agency thereof, and shall not be used for advertising or product endorsement purposes.
This work was in part supported by the National Science Foundation-Department of Energy (DOE) partnership for plasma science and engineering (grant DE-SC0016248), by the DOE-National Nuclear Security Administration (grant DE-NA0003842), and the University of California Laboratory Fees Research Program (grant LFR-17-449059). This work was also performed under the auspices of the U.S. Department of Energy by Lawrence Livermore National Laboratory under Contract No. DE-AC52-07NA27344. This research used computational resources of the National Energy Research Scientific Computing Center, a DOE Office of Science User Facility supported by the Office of Science of the U.S. Department of Energy under Contract No. DE-AC02-05CH11231. 
\end{acknowledgements}


%

\end{document}